\documentclass[a4paper,fleqn,usenatbib]{mnras}

\usepackage{newtxtext,newtxmath}

\usepackage[T1]{fontenc}
\usepackage{ae,aecompl}

\usepackage{graphicx}	
\usepackage{amsmath}	
\usepackage{amssymb}	

\usepackage{bm}

\newcommand{\V}{\bm{v}}

\newcommand{\rp}{r_{\rm p}}

\newcommand{\rH}{r_{\rm H}}

\newcommand{\Omegap}{\Omega_{\rm p}}

\newcommand{\xs}{x_{\rm s}}

\newcommand{\BV}{Brunt-V\"ais\"al\"a\ }

\hypersetup{draft}   

\title[Low-mass planets in three-dimensional discs]{Low-mass planet migration in three dimensional wind-driven inviscid discs: a negative corotation torque}

\author[C.~P.~McNally et al.]{
Colin P.~McNally,$^{1}$\thanks{E-mail: c.mcnally@qmul.ac.uk (CPM)}
Richard P.~Nelson,$^{1}$
Sijme-Jan Paardekooper, $^{1,2}$ \newauthor
Pablo Ben\'itez-Llambay $^{3}$
and Oliver Gressel $^{4,3}$
\\
$^{1}$Astronomy Unit, School of Physics and Astronomy, Queen Mary University of London, London E1 4NS, UK\\
$^{2}$DAMTP, University of Cambridge, Wilberforce Road, Cambridge CB3 0WA, UK\\
$^{3}$ Niels Bohr International Academy, Niels Bohr Institute,
Blegdamsvej 17, DK-2100 Copenhagen \O, Denmark \\
$^{4}$ Leibniz-Institut f\"ur Astrophysik Potsdam (AIP), An der Sternwarte 16, 14482 Potsdam, Germany
}

\date{Accepted: 2020 February 24. Revised: 2020 February 22; Received: 2019 November 22}

\pubyear{2020}

\begin{document}
\label{firstpage}
\pagerange{\pageref{firstpage}--\pageref{lastpage}}
\maketitle

\begin{abstract}
We present simulations of low-mass planet--disc interactions in inviscid three-dimensional discs. 
We show that a wind-driven laminar accretion flow through the surface layers of the disc does not significantly modify the migration torque experienced by embedded planets.
More importantly, we find that 3D effects lead to a dramatic change in the behaviour of the dynamical corotation torque 
compared to earlier 2D theory and simulations. 
Although it was previously shown that the dynamical corotation torque could act to slow and essentially stall the 
inward migration of a low-mass planet, our results in 3D show that the dynamical corotation torque has the 
complete opposite effect and speeds up inward migration.
Our numerical experiments implicate buoyancy resonances as the cause. 
These have two effects: (i) they exert a direct torque on the planet,
whose magnitude relative to the Lindblad torque is measured in our simulations to be small; 
(ii) they torque the gas librating on horseshoe orbits in the corotation region and drive evolution of its vortensity, 
leading to the negative dynamical corotation torque.
This indicates that at low turbulent viscosity, the detailed vertical
thermal structure of the protoplanetary disc plays an important role in determining
the migration behaviour of embedded planets.  If this result holds up under a more refined treatment of disc thermal evolution, then it has important implications for understanding the formation and early evolution of planetary systems.
\end{abstract}




\section{Introduction}

Planet formation occurs in gaseous protoplanetary discs,
and the gravitational interaction between the forming planet and its host disc
plays a central role in determining the evolution and fate of the planetary system.
Due to the combination of very low ionization levels and high optical depths,
 protoplanetary discs, at a few astronomical units from a central solar-type star,
are likely characterized by a largely laminar flow.
In lieu of a vigorous instability able to drive turbulence providing a significant turbulent viscosity,
accretion in these regions of the disc is likely to be driven by a magnetothermal wind, acting in the thin surface layers where
external radiation provides sufficient ionization for magnetic fields to couple to the flow \citep{2013ApJ...769...76B,2015ApJ...801...84G,2016ApJ...818..152B}.
Thus, it is important to understand the interactions between a planet and disc in this context.
In the common technique of modelling the effects of turbulence with a viscous stress,
this scenario corresponds to an inviscid flow.

A planet with a sufficiently low mass will not open a gap in a protoplanetary disc,
but the gravitational effect of the planet will drive features in the disc flow, particularly where the flow of disc
gas past the planet results in a resonant forcing.
The most commonly addressed of these features results from waves excited at the Lindblad resonances, which form a spiral wake.
The first-order Lindblad resonances arise from a coincidence between the differential orbital frequency of the planet and disc gas,
 and the epicyclic frequency of oscillations of the disc gas in the plane of the disc.
The overdensities of the wake structure then exert a tidal gravitational pull on the planet, resulting in a planet migration torque \citep{1980ApJ...241..425G}.
Material in the corotation resonance with the planet can also exchange angular momentum with the planet, resulting in a migration torque \citep{1991LPI....22.1463W}.
Spiral features can also arise due to buoyancy resonances, where primarily vertical oscillations of the disc gas, with restoring force provided by buoyancy forces,
resonate with the differential orbital frequency
of the planet and disc gas \citep{2012ApJ...758L..42Z,2015ApJ...813...88Z}.

These concepts are not purely theoretical.
In the case of embedded planets, the kinematics of the flows induced by planet--disc
interactions are strongly suggested in ALMA observations.
These features are local deviations from Keplerian motion which form part of the Lindblad resonance driven
spiral wake \citep{2018ApJ...860L..13P,2019arXiv190606302C}.
A buoyancy-driven spiral pattern may have been observed in TW~Hya \citep[][Bae in preparation]{2019ApJ...884L..56T}.
If these observations are in fact the kinematic features produced by planet--disc interactions of embedded planets, they would be strong circumstantial evidence
that these planet--disc interactions must be causing angular momentum exchange between the planet and disc, and hence currently driving planet migration in those systems.

Significant effort has been expended on understanding the interaction of low-mass planets
with low-viscosity discs in two dimensional models, and in \citet{2019MNRAS.484..728M} we presented a map of the phenomenological regimes in terms of the disc viscosity and planet mass.
In a region of a protoplanetary disc characterized by a laminar, wind-driven structure,
we expect embedded planets to largely fall in a block
of parameter space where the radial migration behaviour is dominated by dynamical corotation torques
because of the very low effective turbulent viscosity of the disc \citep{2014MNRAS.444.2031P}.
This theoretical understanding of inviscid disc--planet interactions is derived from 1D and 2D models of protoplanetary discs.
In 2D, we have previously modelled low-mass planet--disc interactions with a laminar radial flow
in the mid-plane enabled by a horizontal magnetic field arising from the Hall effect \citep{2017MNRAS.472.1565M,2019MNRAS.484..728M,
2018MNRAS.477.4596M}.
However in 2D it is not possible to directly model the effect of a thin wind-driven accretion layer at the disc surface.
This requires 3D simulations.

Dynamical corotation torques typically arise from a combination of the ability of material trapped on
librating streamlines to conserve its vortensity as the planet migrates, leading to a vortensity contrast developing between librating material and the background disc gas, paired with the geometrical asymmetry in front of and behind the
radially moving planet
 \citep{2006MNRAS.370..784O,2014MNRAS.444.2031P}.
 These theoretical treatments conclude that in discs with radially decreasing vortensity\footnote{Usually corresponding to
 radial surface density power laws flatter than $r^{-3/2}$.}, as a planet migrates inwards,
 the dynamical corotation torque contribution is positive, and acts to slow down planet migration.
 However, in this work, we find 3D configurations with the surprising and opposite outcome,
 that dynamical corotation torque effects accelerate inward planet migration.

Of the simulation studies which have addressed planet--disc interaction in three dimensions,
only very limited set have addressed the low-viscosity regime.
 \citet{2016ApJ...817...19M} studied the torques on low-mass planets
 in globally isothermal inviscid discs, although limited to
 integrations over a timescale of 20 local orbits with a fixed planet.
These models successfully demonstrated a non-linear horseshoe torque acting in
three dimensions at early times after introduction of the fixed planet.
The focus on the short period after the initial introduction of the planet to the flow,
and the fixed planet orbit caused
those simulations to yield qualitatively different results from the ones presented in this work.

Most studies of low-mass planet--disc interaction in three dimensions have considered viscous discs.
Viscous smoothing of the flow suppresses the conditions which lead to the Rossby wave instability forming vortices,
and unsaturates the classical corotation torque,
preventing dynamical corotation torques from
occurring \citep[e.g.][]{2006ApJ...652..730M,2009A&A...506..971K, 2014MNRAS.440..683L,2015ApJ...811..101F,2017AJ....153..124F}.
An important subset of these 3D studies in viscous discs has addressed the torque effects due to
 the release of accretion heating energy from the planet back into the surrounding flow,
 a  phenomena which will not be addressed in this work \citep{2015Natur.520...63B,2019A&A...626A.109C}.

Our previous work has largely been limited to considering globally isothermal gas thermodynamics.
However, the entropy gradients driven by the flows near the planet are an additional driver for
Rossby wave instability, which in the near-adiabatic conditions of the inner disc is
 expected to be virulent \citep{2001MNRAS.326..833B}.
Additionally, in 3D the vertical response of the flow to the planet potential has an important
dependence on the vertical stratification and gas thermodynamics.
The gas buoyancy response has previously been studied as a direct source for migration torques
\citep{2012ApJ...758L..42Z}, including an analytical treatment of the torque due to the buoyancy resonance in a
shearing-sheet approximation \citep{2014ApJ...785...32L}.
A demonstration of the buoyancy response in a global disc model was given by \citet{2015ApJ...813...88Z},
but the topic has otherwise been largely unaddressed.
We will present strong circumstantial evidence that this buoyancy response of the disc
can play a dramatic role in the process that produces dynamical corotation torques.

In this work, we focus on high-resolution, long time-scale simulations including free movement of the planet in a disc model
appropriate for the optically thick inner region of a protoplanetary disc.
These are purely inviscid models, and adopt a planet mass well in the embedded regime.
In three dimensions, we are able to introduce a purely hydrodynamic model for the laminar wind-driven accretion
flow at the surface of the disc.
In this paper, we settle an important outstanding question:
Does a wind-driven laminar accretion flow through the surface layers of the disc significantly modify the migration torque experienced by embedded planets? The answer is no.
We also reveal two phenomena not shown before: 1) The unsurprising presence of vortices in the corotation region. 2) The surprising new role of the buoyancy response of the disc in modifying dynamical corotation torques in 3D.
Section~\ref{sec:theory}  presents a review of previous theoretical work and expectations for the inviscid disc limit.
In Section~\ref{sec:methods} we discuss the physical and numerical aspects of our 2D and 3D simulations, along with defining quantities used in their analysis.
In Section~\ref{sec:results} we proceed by presenting first the results of 2D models, which are largely in agreement with previous 2D expectations, and then 3D models.
Section~\ref{sec:buoyancy} presents more detailed analysis and experiments on the role of the buoyancy response in 3D models,
and Section~\ref{sec:irrad} presents a brief test of an alternate model appropriate for a passively irradiated disc.
Finally, we discuss our results and their implications for planet migration in a broader context in Section~\ref{sec:discussion},
and present our conclusions in Section~\ref{sec:conclusions}.

\begin{table*}

\caption{Main simulation models discussed in this work.}
\label{tab:runs}
\begin{tabular}{lcclcc}
\hline
Model & Dimensionality & Gas properties & Vertical structure & Planet properties & Resolution \\
\hline
2AMS & 2D & adiabatic (no thermal relaxation) & isothermal  $h\propto r$ &  moving planet  & single resolution \\
2AMD & 2D & adiabatic (no thermal relaxation)  & isothermal  $h\propto r$ & moving planet & double resolution \\
2AFS & 2D & adiabatic  (no thermal relaxation) & isothermal  $h\propto r$ & fixed planet & single resolution \\
3WMS & 3D & thermal relaxation + wind driving & isothermal  $h\propto r$ & moving planet & single resolution \\
3AMS & 3D &adiabatic (no thermal relaxation)  & isothermal  $h\propto r$ & moving planet & single resolution \\
3AFS & 3D &adiabatic  (no thermal relaxation) & isothermal  $h\propto r$ & fixed planet & single resolution \\
3AMH & 3D &adiabatic  (no thermal relaxation) & isothermal  $h\propto r$ &moving planet  & half resolution \\
3AFSP & 3D & adiabatic  (no thermal relaxation) & polytropic  $h\propto r$ & fixed planet & single resolution \\
B2AMS & 2D & adiabatic (no thermal relaxation)  & isothermal  $h\propto r^{2/7}$ &  moving planet  & single resolution \\
B3WMS & 3D & thermal relaxation + wind driving & isothermal  $h\propto r^{2/7}$ & moving planet & single resolution \\
B3AFS & 3D & adiabatic  (no thermal relaxation) & isothermal $h\propto r^{2/7}$ & fixed planet & single resolution \\
3AFSV & 3D & adiabatic  (no thermal relaxation) + viscous & isothermal $h\propto r$ & fixed planet & single resolution \\
3AFSPV & 3D & adiabatic  (no thermal relaxation) + viscous & polytropic $h\propto r$ & fixed planet & single resolution \\
\hline
\end{tabular}
\end{table*}

\section{Theoretical expectations}
\label{sec:theory}
Our previous work has examined the migration of low-mass planets in discs with low or zero turbulent viscosity, in both the absence and presence of laminar accretion flows generated by large-scale magnetic fields \citep{2014MNRAS.444.2031P, 2017MNRAS.472.1565M, 2018MNRAS.477.4596M}. These studies were conducted for 2D vertically integrated and globally isothermal disc models. In the presence of a laminar accretion flow the models apply specifically to the case where the magnetic field torques the gas, and induces a radial flow, near the disc mid-plane.

The migration rate is determined by the instantaneous contributions from Lindblad and corotation torques. Working in the limit of zero viscosity and a simple power-law disc model, we expect the Lindblad torque to have a well-defined negative value that is determined by the surface density and temperature profiles in the gas. The corotation torque, however, is expected to evolve according to the following expression as the planet migrates \citep{2017MNRAS.472.1565M}
\begin{align}
\Gamma_{\rm hs} = 2\pi \left( 1-\frac{\omega(\rp)}{\omega_{\rm c} (t)}\right) \Sigma_{\rm p} \rp ^2 x_s \Omega_{\rm p} \left[\frac{{\rm d} \rp}{{\rm d}t} -v_r \right],
\label{eq:unified}
\end{align}
and hence determines the long-term migration behaviour. The symbols have the following meanings: $\omega_{\rm c}(t)$ is the characteristic vortensity of material on librating streamlines trapped on horseshoe orbits,
$\omega(\rp)$ is the vortensity of the unperturbed background disc at the planet location,
$\Sigma_{\rm p}$ is the disc surface density at the planet position,
$\rp$ is the planet's radial position (such that $d \rp/dt$ is the migration speed),
$x_s$ is the half-width of the corotation region,
$\Omega_{\rm p}$ is the Keplerian orbital frequency at the planet position,
and $v_r$ is the radial velocity of the disc gas flow. Note that negative values of ${\rm d} \rp/{\rm d}t$ and $v_r$ indicate migration and gas flow towards the star.
In a 2D disc, the vortensity can be written as $\omega= (\nabla \times \V)/\Sigma$.

In \citet{2017MNRAS.472.1565M, 2018MNRAS.477.4596M} we identified and demonstrated the existence of four regimes of behaviour in the above torque expression~(\ref{eq:unified}), separated by sign changes in two quantities related to the planet migration rate, ${\rm d}r_{\rm p}/ {\rm d}t$, and the gas flow speed, $v_r$. These regimes correspond to \emph{qualitative differences} in the expected migration behaviour of a planet. If we restrict our discussion to Keplerian discs with radially decreasing vortensity profiles, and for which the radial flow of gas is directed towards the star and the planet naturally wants to migrate inwards, then these four regimes of behaviour reduce to two possible regimes. Since vorticity $\omega=\Omega/(2\Sigma)$, where $\Omega$ is the Keplerian orbital angular velocity, a power-law disc model with a radially decreasing vortensity has $\Sigma \propto r^{-\delta}$ where $\delta\le3/2$.

Before we discuss these two regimes, it is important to note that
the gas trapped on horseshoe orbits in the planet's corotation region moves with the planet as it migrates. 
In the absence of any magnetic torque acting on it, the vortensity of this gas, $\omega_{\rm c}$, is conserved
\citep{2014MNRAS.444.2031P, 2017MNRAS.472.1565M}.
If a magnetic torque operates and drives an inward gas flow, the vorticity, $\omega_{\rm c}$, will increase with time. Equation~\ref{eq:unified} shows that the corotation torque depends on how the ratio $\omega(\rp)/\omega_{\rm c} (t)$ evolves. For an inward migrating planet in a disc with a radially decreasing vortensity profile, $\omega(\rp)$ must increase as the planet moves inwards. The rate at which $\omega_{\rm c} (t)$ evolves then has a controlling influence on how the dynamical corotation torque evolves.

The parameters determining the borders between migration regimes are the disc gas radial flow velocity $v_r$ and the `initial' planet migration velocity ${\rm d} \rp/{\rm d} t$, which is the value obtained when considering the effects of Lindblad torques only.
Given this, and the above discussion about the evolution of $\omega_{\rm c}$, the two regimes of interest, and the expected long-term migration behaviours predicted by equation~(\ref{eq:unified}), are:
\begin{description}
\item[ (i) $v_r \leq  0$ and {$[{\rm d} \rp/{\rm d}t - v_r] <0$}:] 
The disc accretion flow and planet migration are inwards, and the planet initially migrates faster than the disc flow, such that $\omega(\rp)$ increases faster than $\omega_{\rm c}(t)$ $\implies$ \emph{The planet migrates inwards, and asymptotically approaches the gas inflow speed}. In the limit of $v_r \rightarrow 0$, we see that the value of $\omega_{\rm c}$ remains constant and $\omega(\rp)$ increases in equation~(\ref{eq:unified}), such that $\Gamma_{\rm hs}$ increases with time. Planet migration thus slows down and eventually stalls completely.
\item[ (ii) $v_r < 0$ and {$[{\rm d} \rp/{\rm d}t - v_r]>0$}:]
The disc accretion flow and planet migration are inwards, and the planet initially migrates slower than the disc flow $\implies$ \emph{The planet's inward migration slows down, reverses, and then runs away outwards, producing very rapid outward migration}.
 \end{description}

The simulations we present in this paper are for 3D disc models, with and without laminar accretion flows. Our expectation for the models without radial accretion flows is that they will show behaviour described by regime (i) with $v_r=0$, namely the slowing of migration as the planet migrates inwards.
The introduction of a  more complete description of disc thermodynamics ought to cause the baroclinic forcing of vortices at the edges of the horseshoe region, due to the radial mixing of entropy. These might have the effect of partially mixing the vortensity in the horseshoe region with the surrounding disc, weakening the corotation torque. This would cause migration to not stall completely, but to continue inwards at some finite rate which is slower than that induced by Lindblad torques alone.

The simulations performed with accretion flows include a torque that drives a laminar flow in the surface layers of the disc, to mimic the torque expected when a magnetized wind is launched under the assumption of non-ideal MHD \citep{2013ApJ...769...76B, 2015ApJ...801...84G,2017ApJ...845...75B,2017A&A...600A..75B}. The accretion flow in these simulations is faster than the initial planet migration speed, and hence these simulations provide a test of whether or not the evolution corresponding to regime (ii) above is obtained when the accretion flow is confined to the surface layers rather than being located near the disc mid-plane. Our expectation is that the migration will not be strongly affected as the accretion flow occurs at high altitudes in the disc where there is little mass present.

Finally, one consequence of our more refined treatment of disc thermodynamics is that buoyancy resonances may provide an additional torque on the planet,  with the inner/outer buoyancy resonances augmenting the magnitudes of the inner/outer Lindblad torques by $\sim 10\%$ 
for disc and planet models like the ones we consider
\citep{2012ApJ...758L..42Z, 2014ApJ...785...32L}. What is not clear is how the torque exerted by the \emph{planet on the disc} will affect the long-term evolution. This is because buoyancy resonances occur radially close to the planet in the corotation region, and \citet{2014ApJ...785...32L} suggest the disc response at the buoyancy resonances is non-wavelike, such that the torque provided by the planet is deposited locally in the disc. This will cause evolution of the vortensity in the coorbital region,  and hence will affect the corotation torque acting on the planet. The magnitude and sign of this effect are unknown, and this is perhaps the most important effect to be unveiled by our simulations.


\begin{figure}
\includegraphics[width=0.9\columnwidth]{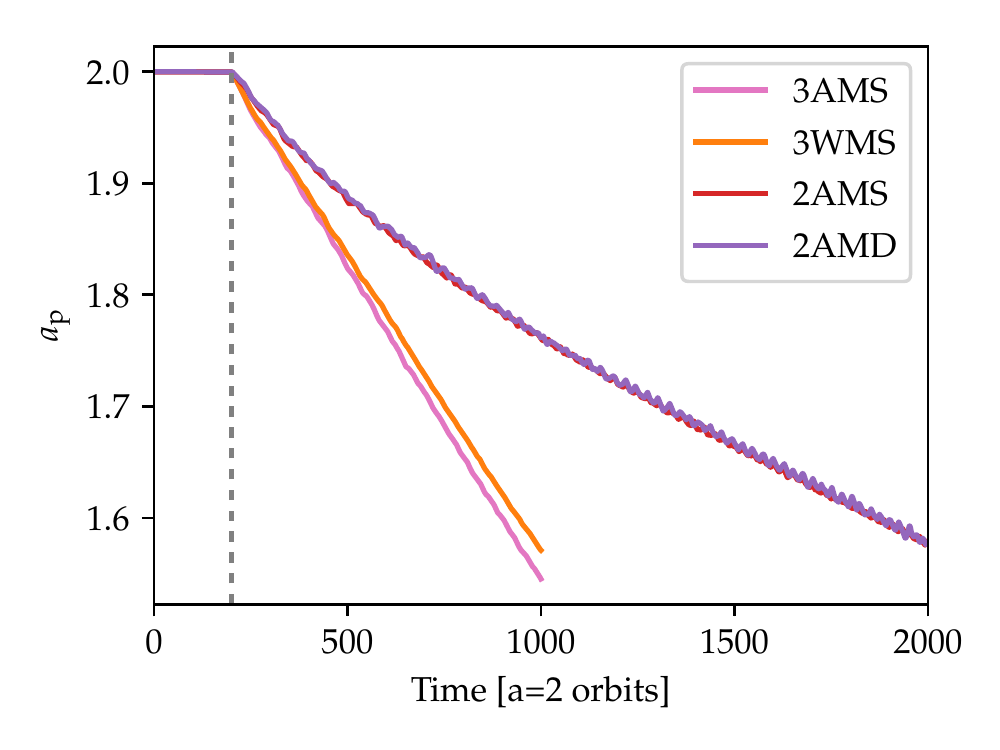}\\
\caption{
Planet migration trajectories in 2D and 3D models. {\sl Grey dashed line:} Planet release time.
Run labels defined in Table~\ref{tab:runs}. Notably, in 3D simulations the planet migrates inwards more
rapidly than in the equivalent 2D simulations.
}
\label{fig:trajectories_2d3d}
\end{figure}

\begin{figure*}
\includegraphics[width=1.8\columnwidth]{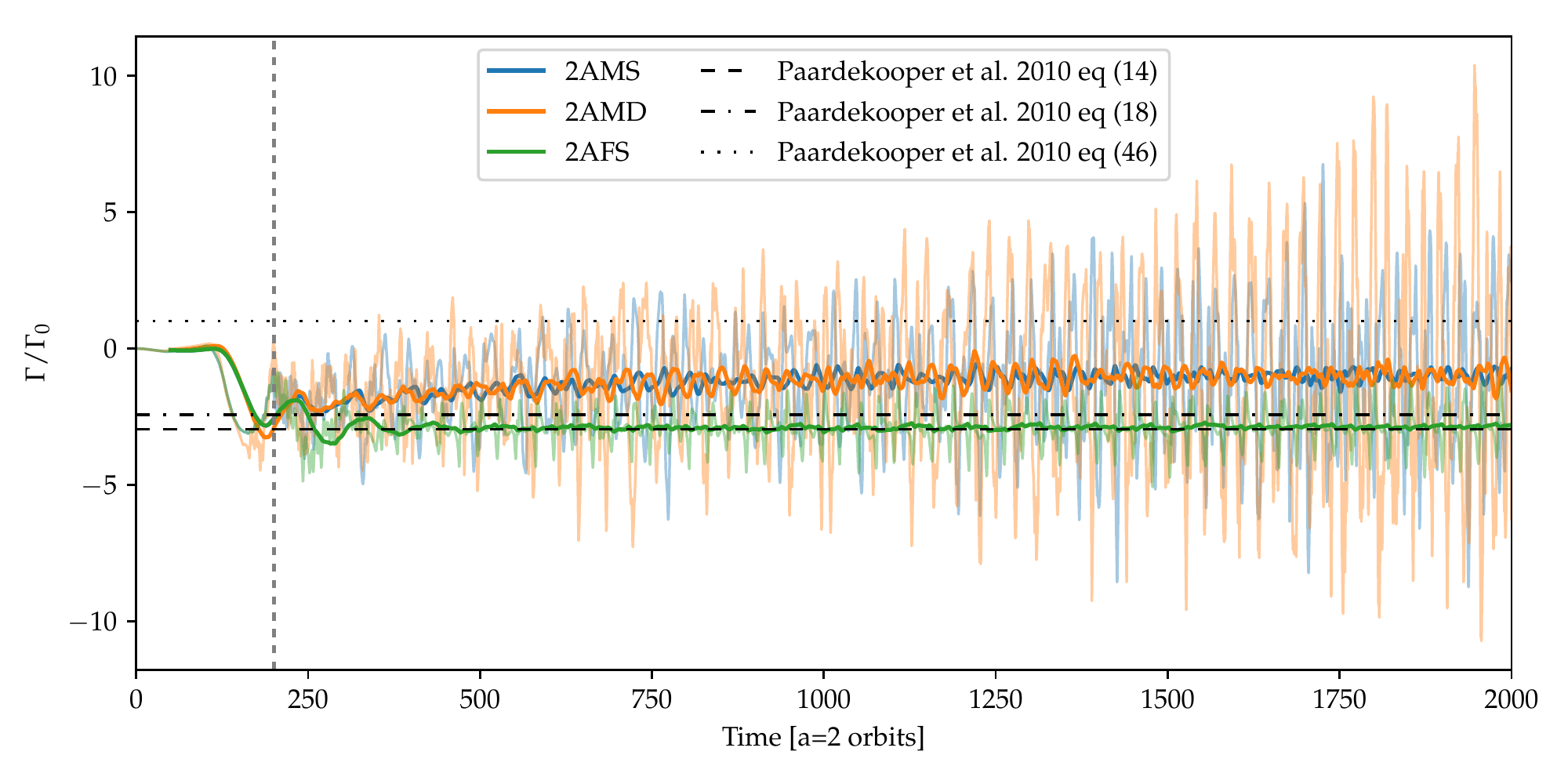}\\
\caption{
Torque histories from 2D simulations. Run names given in Table~\ref{tab:runs}.
{\sl Light lines:} Instantaneous values.
{\sl Solid lines:} 50 orbit trailing averages.
{\sl Grey dashed vertical line:} End of planet mass ramping and planet release.
The single- and double-resolution moving planet cases 2AMS and 2AMD show a positive dynamical corotation torque effect,
while the fixed planet case 2AFS shows excellent agreement with the Lindblad torque formula given by \citet{2008A&A...485..877P}
indicating a saturation of the  corotation torque.
Also plotted are 2D torque formulas from \citet{2010MNRAS.401.1950P}. Their equation~(14) is an analytical estimate of the Lindblad torque, 
equation~(18) is the Lindblad torque and linear corotation torque, and equation~(46) the Lindblad torque and unsaturated non-linear horseshoe corotation torque.}
\label{fig:torques_2AMS_2AMD_2AFS}
\end{figure*}

\begin{figure}
\includegraphics[width=\columnwidth]{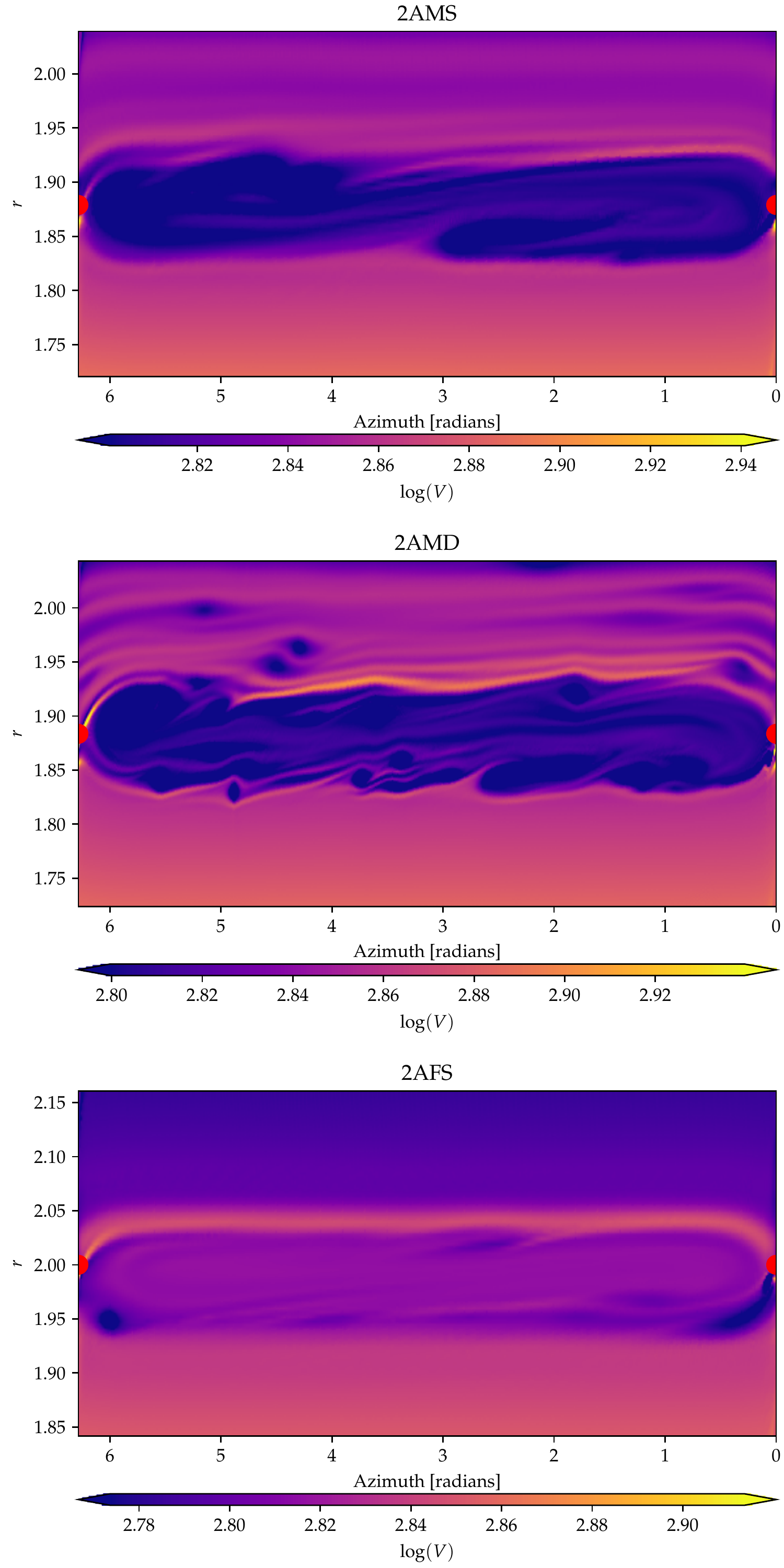} \\
\caption{
Vortensity $V$ at 500 orbits in the two-dimensional runs 2AMS, 2AMD, 2AFS in a strip centred on the planet location.
The colour scales are a fixed logarithmic range about the vortensity of the initial condition at the planet's present location.
The planet location is marked by the red circle split across on the periodic azimuthal boundary.
Videos available on Zenodo archive at {doi:10.5281/zenodo.3613755}
}
\label{fig:V_map_2AMS_2AMD_2AFS}
\end{figure}

\begin{figure}
\includegraphics[width=\columnwidth]{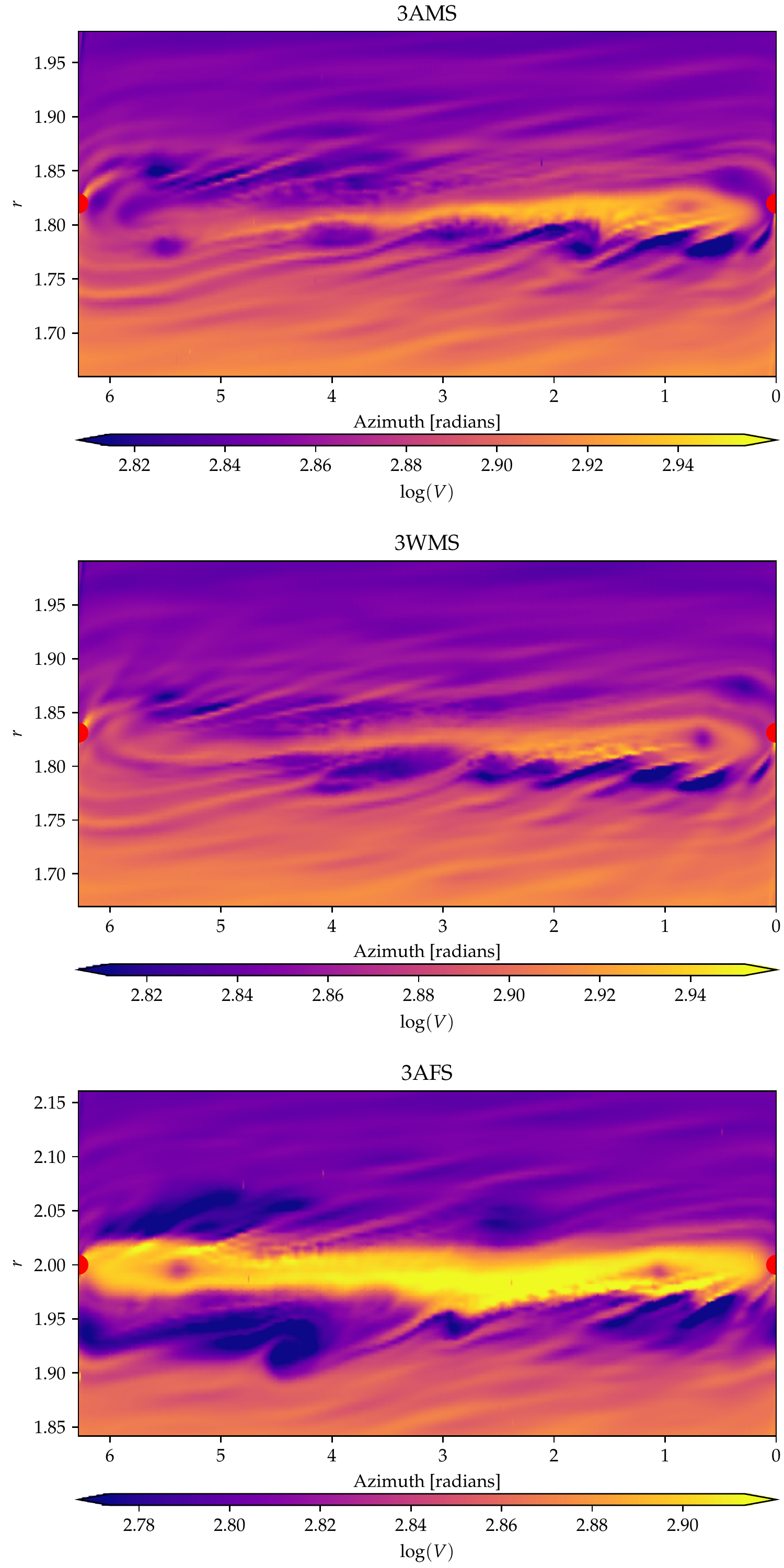} \\
\caption{
Vertically integrated vortensity $V$ at 500 orbits in the three-dimensional
runs 3AMS, 3WMS, 3AFS in a strip centred on the planet location.
The colour scales are a fixed logarithmic range about the vortensity of the initial condition at the planet's present location.
The planet location is marked by the red circle split across on the periodic azimuthal boundary.
Videos available on Zenodo archive at { doi:10.5281/zenodo.3613755}
}
\label{fig:V_map_3AFS_3AMS_3AFS}
\end{figure}

\begin{figure}
\includegraphics[width=0.9\columnwidth]{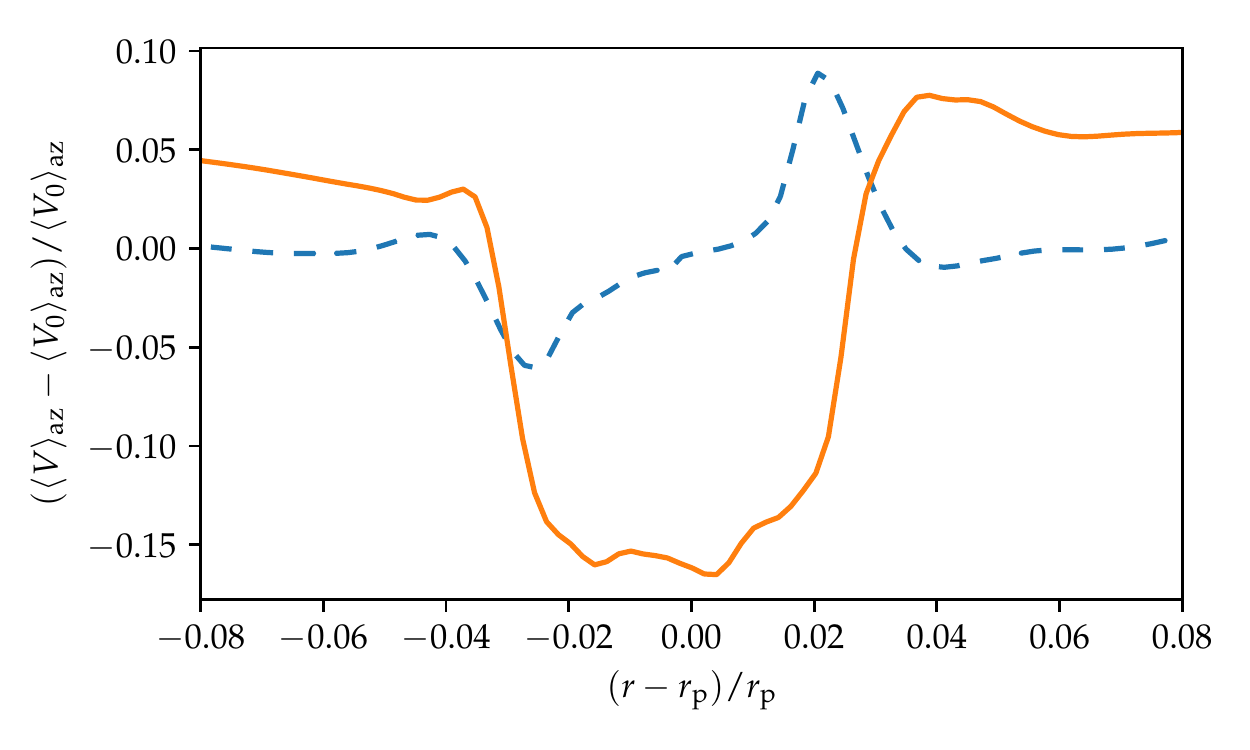} \\
\caption{
Azimuthally averaged vortensity relative perturbation, at 1000 orbits.
This shows that, in 2D, the relative vortensity deficit in the coorbital region occurs only in the moving planet case.
{\sl Dashed line:} 2D fixed planet model 2AFS.
{\sl Solid line:}  2D moving planet model 2AMS.
}
\label{fig:Vavg_2AFS_2AMS}
\end{figure}

\begin{figure*}
\includegraphics[width=1.8\columnwidth]{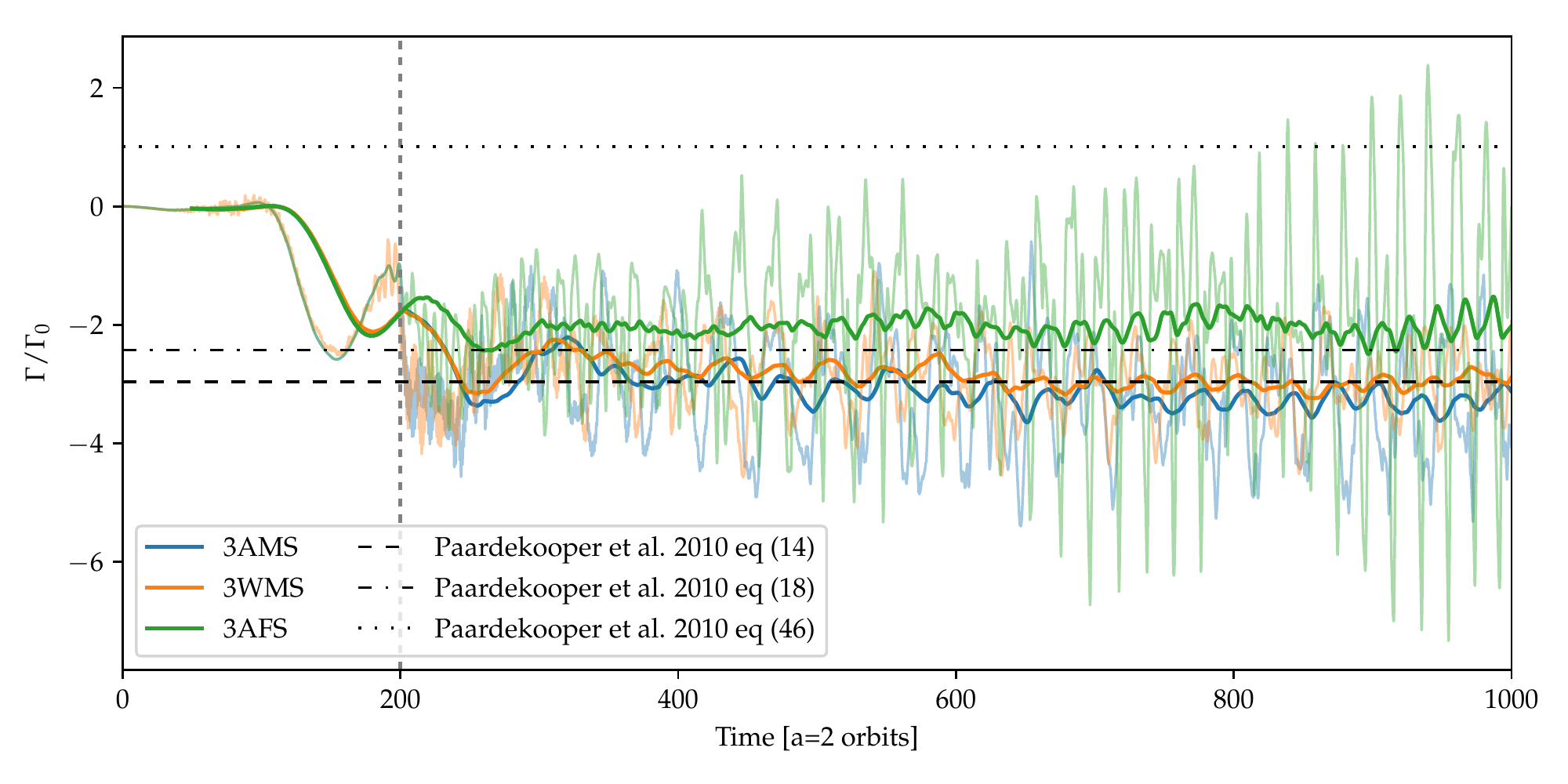}\\
\caption{
Torque histories from 3D simulations. Run names given in Table~\ref{tab:runs}.
{\sl Light lines:} Instantaneous values.
{\sl Solid lines:} 50 orbit trailing averages.
{\sl Grey dashed vertical line:} End of planet mass ramping and planet release.
The moving planet adiabatic and wind-thermal-relaxation cases 3AMS and 3WMS are not significantly distinguishable,
while both show a more negative torque than the fixed planet case 3AFS, indicating a negative dynamical torque effect.
Also plotted are 2D torque formulas from \citet{2010MNRAS.401.1950P}. Their equation~(14) is an analytical estimate of the Lindblad torque, 
equation~(18) is the Lindblad torque and linear corotation torque, and equation~(46) the Lindblad torque and unsaturated non-linear horseshoe corotation torque.}
\label{fig:torques_3AMS_3WMS_3AFS}
\end{figure*}

\begin{figure}
\includegraphics[width=\columnwidth]{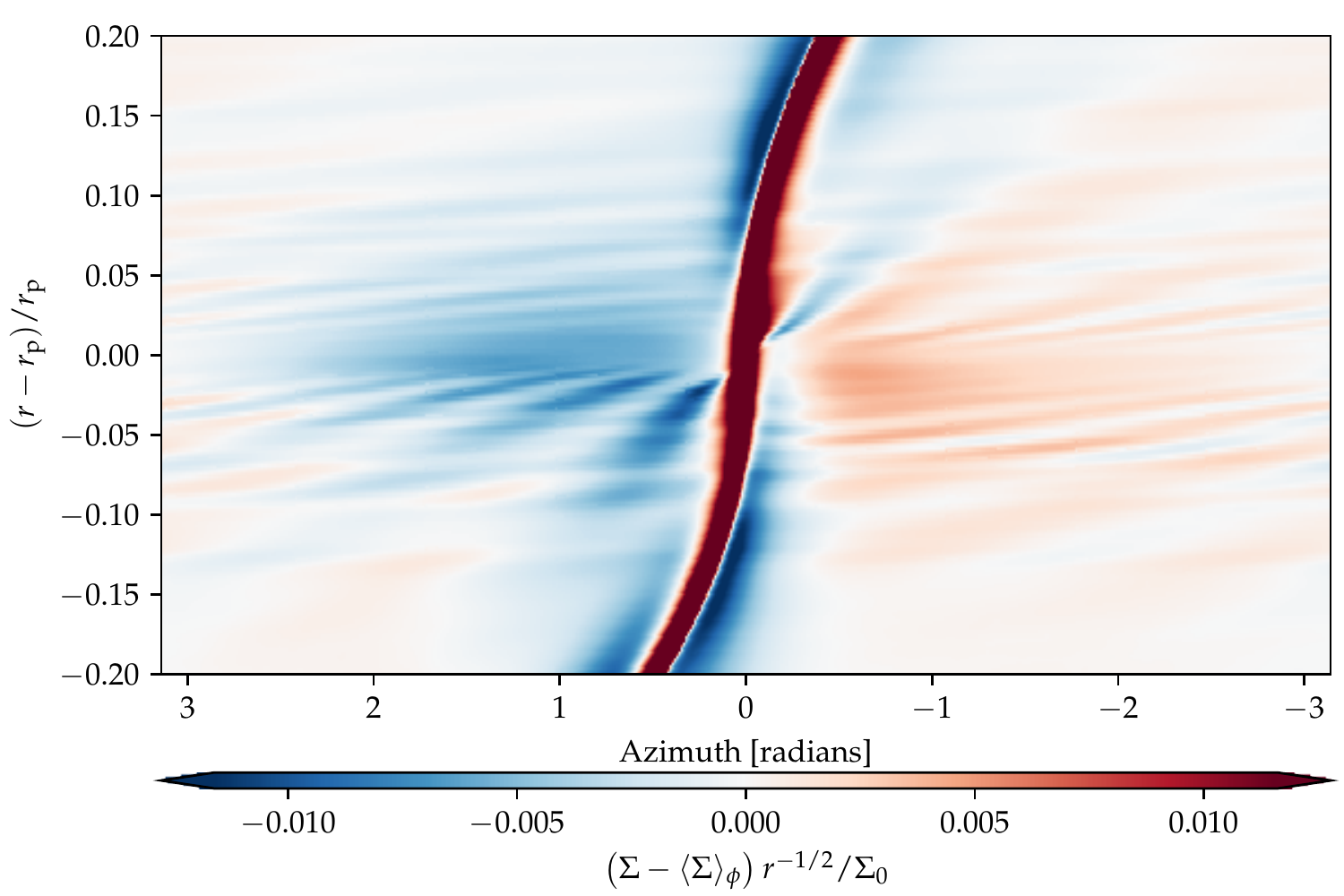}\\
\includegraphics[width=\columnwidth]{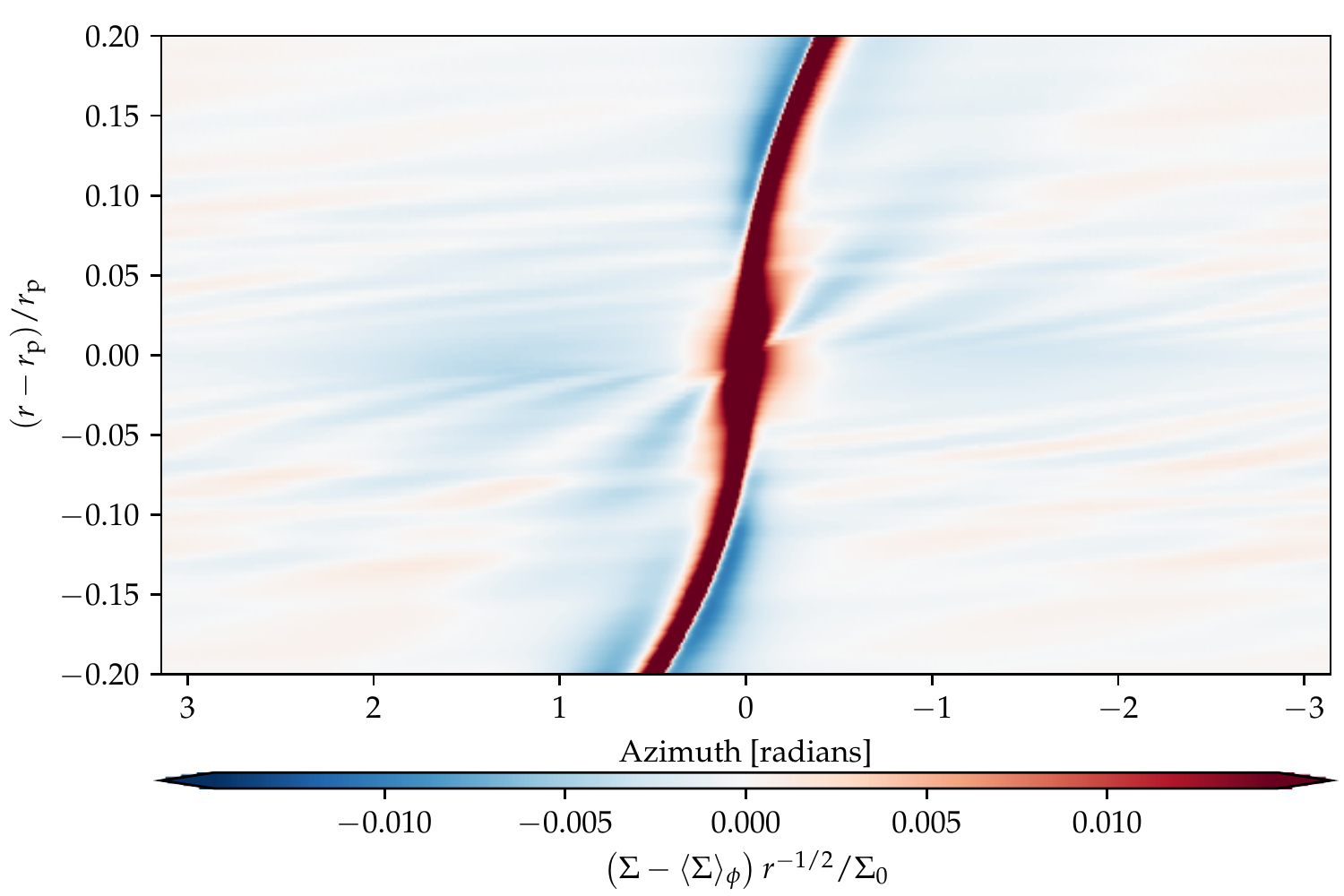}
\caption{
Time-averaged azimuthal surface density perturbations averaged from 600 to 700 orbits.
{\sl Top:} Moving planet run 3AMS.
{\sl Bottom:} Bottom fixed run 3AFS.
In both cases the planet is located at the origin of the coordinate system.
The moving planet case shows an azimuthal asymmetry not present in the fixed planet case,
with a density enhancement in front of the planet in its orbit, to the right on the upper panel.
In contrast, the density contrast distribution is more symmetric in the fixed planet case (lower panel).
Also visible are rays of overdensity and underdensity to the upper left, and lower right of the planet, originating from the buoyancy response.
}
\label{fig:coldensdiff}
\end{figure}

\begin{figure}
\includegraphics[width=0.9\columnwidth]{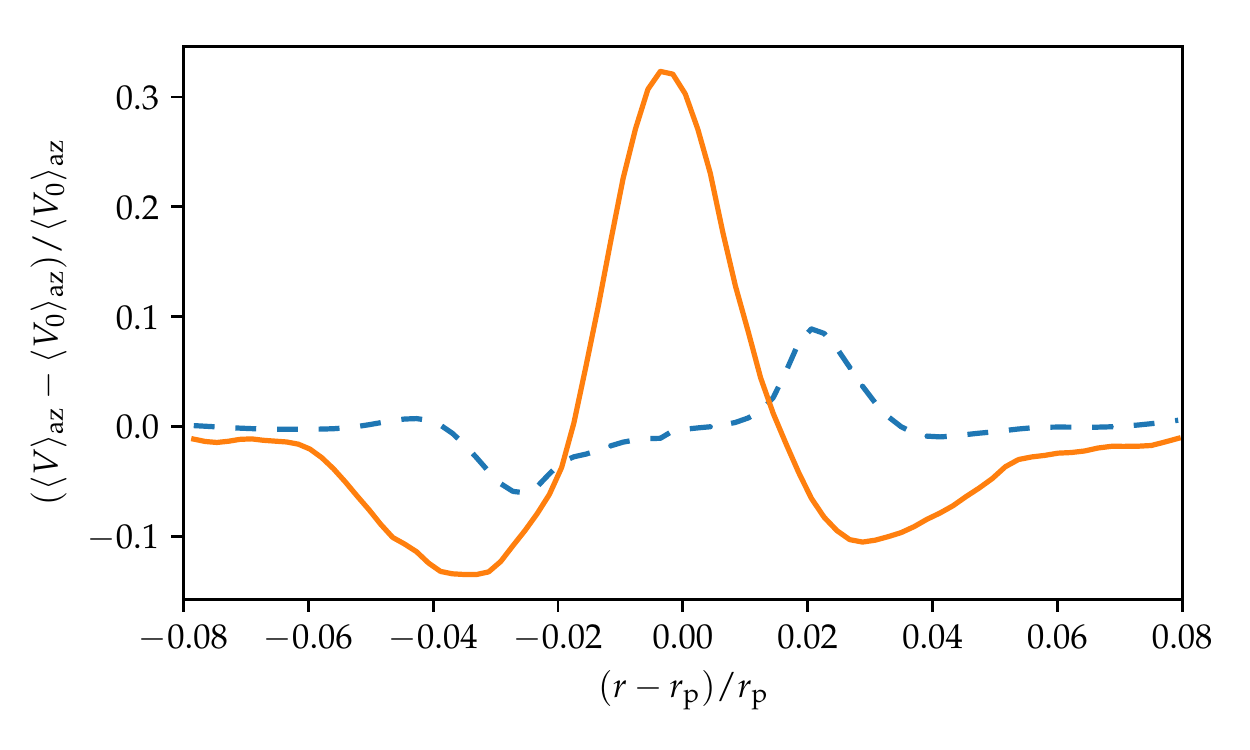} \\
\caption{
Azimuthally averaged vertically integrated vortensity relative perturbation, at 1000 orbits.
This shows that the vortensity enhancement in the coorbital region occurs only in the three-dimensional model.
{\sl Dashed line:} Two-dimensional model 2AFS.
{\sl Solid line:} Three-dimensional model 3AFS.
}
\label{fig:Vavg_2AFS_3AFS}
\end{figure}

\section{Methods}
\label{sec:methods}

The goal of the models considered in this work is to capture the conditions typical of the inner dead zone of a protoplanetary disc,
at stellocentric radii of a few au.
In this region the bulk of the disc column has very high magnetic resistivity,
so that the magnetorotational instability is impeded.
The disc thermodynamics are very close to adiabatic, with long cooling
 times in the first few scale heights away from the mid-plane.
 As the disc is both very optically thick to its own thermal radiation, and is heated from the surfaces,
  not from internal viscous dissipation, its temperature structure is vertically isothermal.

The simulations presented here solve the compressible Euler equations for gas dynamics with an ideal gas equation of state.
In three dimensions the basic form of these are:
\begin{align}
\frac{\partial \rho}{\partial t} &= -\nabla \cdot (\rho \V) \label{eq:continuity}\\
\frac{\partial{\V}}{\partial t}  &= - (\V \cdot \nabla) \V  + \Omega_{\rm F}^2 r \hat{\bm e}_r - 2\Omega_{\rm F}\times \V -\nabla \Phi - \frac{\nabla P}{\rho}\\
\frac{\partial s}{\partial t} &= - (\V \cdot \nabla) s + \left(\frac{\partial s}{\partial t}\right)_{\rm relax} \label{eq:dsdt}\\
P &= (\gamma-1) \rho c_{\rm v} T \label{eq:eos},
\end{align}
where
$\rho$ is the gas volume density,
$\V$ is the gas velocity in the frame rotating with angular frequency $\Omega_{\rm F}$,
$\Phi$ is the gravitational potential (of both the central star and planet),
$s$ is the gas specific entropy,
$P$ is the gas pressure,
$\gamma$ is the adiabatic index (taken as 1.4),
$c_{\rm v}$ is the gas specific heat capacity, and
$T$ is the gas temperature. The last term in equation~(\ref{eq:dsdt}) represents a thermal relaxation term that is described below.
Excepting a pair of specialized runs presented in Section~\ref{sec:torquemapbuoyancy}, all the simulations consider inviscid gas.
For two-dimensional simulations, the vertically integrated version of these equations is used, with the gas surface density being denoted $\Sigma$.
Specific entropy can be expressed as
\begin{align}
s \equiv c_{\rm v} \log\left( \frac{T}{T_0}\left(\frac{\rho}{\rho_0}\right)^{-(\gamma-1)}\right),
\end{align}
where $\rho_0$, $T_0$ are arbitrary constants.
The specific heat capacity at constant volume $c_{\rm v}$ is
\begin{align}
c_{\rm v} = \frac{k_{\rm B}}{\mu m_{\rm H}(\gamma-1)},
\end{align}
where $k_{\rm B}$ is the Boltzmann constant, $\mu$ is the mean molecular mass (taken as $2.33$) in terms of the hydrogen mass,
and $m_{\rm H}$ is the mass of a hydrogen atom.
Simulations were run with a
modified version of {\sc FARGO3D}~1.2 \citep{2016ApJS..223...11B}, including hybrid MPI/OpenMP parallelism
and MPI-IO based parallel input-output routines.
The numerical implementation of the specific entropy version of the energy equation is described in \citet{2019MNRAS.489L..17M}. 
Throughout, the standard FARGO3D shock-capturing von Neumann-Richtmyer artificial viscosity (quadratic in divergence of velocity), including momentum and heating terms, is employed.

The disc surface density follows a power law $\Sigma = \Sigma_0 (r/r_0)^{-\alpha}$ with $\alpha = 0.5$,
and $\Sigma_0 = 3.8\times 10^{-4}\ \mathrm{M_\odot \ au^{-2}}$.
Scaled to $r_0=1 \  \mathrm{au}$ this surface density corresponds to twice the Minimum Mass Solar Nebula surface density at $r=1$,
but with a flatter surface density profile, providing a radial vortensity gradient in the vertically integrated disc.

The radial temperature gradient is selected to produce a constant disc aspect ratio $h = H/r = 0.05$.
The radial scaling of the disc aspect ratio is $h \propto r^f$, with flaring index $f=0$.
This corresponds to a radial scaling of temperature $T \propto r^{-\beta}$ with $\beta=1$ and $\beta = 1-2f$.
The radial scaling of specific entropy, in a two-dimensional, vertically integrated disc model,
 is thus $s\propto r^{-\xi}$ with $\xi = \beta-(\gamma-1)\alpha$.
Taking $\gamma=1.4$, this gives $\xi=0.8$, corresponding to a radially decreasing specific entropy gradient in a vertically integrated model.

Unless otherwise specified, we adopt units such that $G=1$, $M_\odot = 1$, and $1\ \mathrm{au}$ is the unit of length.
This makes the Keplerian orbital frequency $\Omega_{\rm K}=1$ at $r=1$.
The grid follows spherical polar coordinates, with $(\text{spherical\ radius},\text{\ azimuthal\ angle},\text{\ polar\ angle})$ $= (r,\phi,\theta)$,
spaced evenly in azimuth and polar angle, and logarithmically in radius $r$ over the interval $[0.8,4]$ and five scale heights above the disc mid-plane
in the polar angle $\theta$ interval $[\pi/2-\arctan(5h), \pi/2]$.
To produce a resolution of 25 zones per scale height, the grid has a resolution $(805,3141,125)$ in the basic case.

\subsection{Boundary conditions}
Symmetry conditions are applied at the mid-plane as we only evolve the upper hemisphere of the disc.
Radial and polar boundary conditions at the edges of the grid for the upper half-disc domain follow
those specified by \citet{2016ApJ...817...19M}.
The base boundary conditions are supplemented by
wave damping zones at the top and radial sides of the domain,
as specified in \citet{2019MNRAS.484..728M}, extended vertically at constant cylindrical radius from the mid-plane.
In the wave killing zones, the region near the upper boundary, and in  relaxation of the initial condition a scalar artificial linear
viscous pressure as in \citet{1992ApJS...80..753S}, their equation~(38), is used.
This is available in {\sc FARGO3D}~1.2 with the macro {\tt STRONG\_SHOCK} in sub-step 2a.
Within one-half scale height of the polar boundary in the disc atmosphere, an artificial linear viscous pressure is applied with a linear windowing function
\begin{align}
W_\theta(\theta) =
\begin{cases}
\left(1- \left(\frac{\theta - \arctan(5 h)}{\arctan(h/2)} \right) \right) \ &\text{if } \theta < \frac{\pi}{2} -\arctan(9h/2) \\
0 &\text{ otherwise}
\end{cases}
\end{align}
to prevent wave reflection from the boundary.
The radial boundary wave killing zones also apply the same scalar artificial linear viscous pressure, windowed
linearly over the same radial range in cylindrical radius, and we refer to that windowing function as $W_r(r)$.
During the initial relaxation of the disc to a numerical hydrostatic equilibrium, this term provides additional damping which speeds up the relaxation, 
and during the rest of the simulation its application provides an additional precaution preventing 
spurious wave reflection from the grid boundaries, which might otherwise corrupt the solution in the body of the domain.

\subsection{Disc initial conditions}
The initial three-dimensional density and velocity distribution for a vertically isothermal disc is given by \citet{2016ApJ...817...19M}.
However, this initial analytical equilibrium is not an exact equilibrium once discretized, and due to the low viscosity and low planet masses
a very quiet initial condition is needed for successful simulations.
Thus, we initialize the azimuthal velocity using exact numerical force balance in the spherical-radial direction along each grid line,
and then relax the entire volume with a bulk viscosity ramped down over time for 50 orbits, before inserting the planet potential at time $t=0$.
This initial relaxation of the axisymmetric initial condition is performed in two-dimensional radial-polar coordinates.

The initial condition is relaxed using an additional windowing function on the artificial linear viscous pressure, with
\begin{align}
W_t(t) &=
\begin{cases}
 \left(1- \left(t/50\text{ orbits} \right)^2 \right) &\text{ if } t < 0\text{ orbits}\\
0 &\text{otherwise}
\end{cases}
\end{align}
making the total windowing applied to the artificial linear viscous pressure
\begin{align}
W(t,\theta) &= \max( W_t(t), W_r(r), W_\theta(\theta)).
\end{align}
Thus, at times $t>0$, only the wave-damping regions near the boundaries have artificial linear viscous pressure applied.

\subsection{Radiative cooling/heating approximation}

We employ temperature forcing as used by \citet{2016ApJ...817..102L} and similar to that commonly employed in planetary
atmosphere modelling to relax the disc towards the initial condition temperature field.
The specific entropy equation is:
\begin{align}
\frac{\partial s}{\partial t} = - \V \cdot \nabla s + \frac{1}{T}\left[ -c_{\rm v} \frac{T-T_{\rm ref}}{t_{\rm relax}}\right],
\end{align}
where the last term is the temperature forcing, $c_{\rm v}$ the the specific heat capacity at constant volume, $T$ the the gas temperature,
$T_{\rm ref}$ is the reference temperature 
implied by consistency with the specified initial gas scale height,
and $t_{\rm relax}$ is the time-scale of temperature relaxation.

The time-scale for temperature relaxation is taken from considering the time-scale for radiative loss of energy from a gaussian sphere of scale height $H$:
\begin{align}
t_{\rm relax} = \frac{c_{\rm v} H \rho \tau_{\rm eff}}{3 \sigma_{\rm SB} T^3},
\end{align}
where $\sigma_{\rm SB}$ is the Stefan--Boltzmann constant, $\tau_{\rm eff}$ is the effective optical depth
\begin{align}
\tau_{\rm eff} &= \frac{3\tau_{\rm R}}{8} + \frac{1}{2} + \frac{1}{4 \tau_{\rm P}},\\
\tau_{\rm R} &= \int^{\infty}_{z_{\rm c}} \kappa_{\rm R} \rho dz, \label{eq:tau_integral},
\end{align}
which is the appropriate form of the approximation of \citet{1990ApJ...351..632H} for an irradiated disc \citep{2012ApJ...757...50D},
and we approximate that the Rosseland and Planck mean opacities ($\kappa_{\rm R}$, $\kappa_{\rm P}$) are equal.
The integral equation~(\ref{eq:tau_integral}) is approximated by summing along columns in the meridional direction of the spherical grid.

\subsection{Wind driving}
In some simulations,
to hydrodynamically mimic the effect of a magnetocentrifugal wind driving a radially inward accretion flow across the surface of the disc in what
would be magnetically coupled layers activated by stellar UV radiation, we include a forcing term similar to that employed in
\citep{2017MNRAS.472.1565M,2018MNRAS.477.4596M} which produces a radial inflow. Here it is applied to a constant column density from
the surface of the disc, so as to produce a radially constant mass flux.
After integrating the column density $\Sigma_{\rm z}$ on spherical grid columns down from the upper boundary, the driving torque is
windowed with the function\footnote{This approach has also been used to study wind-driven discs with embedded high-mass planets in collaboration with Elena Lega, Alessandro Morbidelli and Aur\'elien Crida (in preparation).} $W_{\rm w}(\Sigma_{\rm z})$, a Fermi distribution:
\begin{align}
W_{\rm w}(\Sigma_{\rm z}) &= \left[\exp\left(\frac{\Sigma_{\rm z}/\Sigma_{\rm w}-1.0}{D}\right) +1 \right]^{-1},
\end{align}
where $\Sigma_{\rm w}$ is the total column density of surface material driven by the torque and the transition smoothing parameter is taken as $D=0.1$.

Where wind driving is used, the wind is chosen to be `maximal' in the sense that it penetrates deeper into the disc than any expected physical wind.
The driven surface density is set, in physical units, to $\Sigma_{\rm w}  = 10^{-5}\ \mathrm{M_\odot\ au^{-2}}$, and
the torque to produce a wind velocity at $r_0=1\ \mathrm{au}$ of $v_{r0} = -6.32305\times10^{-5}\ \mathrm{au\ \Omega_0/(2\pi)}$
producing in total of the two sides of the disc an accretion rate of $\dot{M} = 8\times10^{-7} \mathrm{M_\odot\ yr^{-1}}$.
This is very high in comparison to the accretion rates onto typical Class-II T-Tauri stars, so should serve as a limiting case for
possible influence of the wind.
The wind advection moves low-density surface material inwards, to regions where the disc atmosphere would in equilibrium be hotter.
As the physical cooling time of this gas is short, to maintain physical consistency and prevent an unphysical thermal inversion from developing,
the thermal relaxation scheme is always employed with the wind driving.

\subsection{Planet Potential}
The planet mass to central stellar mass ratio was $q=2\times10^{-5}$ corresponding to $6.7\ \mathrm{M_\oplus}$ planet in orbit at 2 au
around a solar mass star.
At this radial position, the Keplerian orbital frequency $\Omega_{\rm p}=0.3536$ in code units, and the adiabatic sound speed is $c_{\rm s}=0.0418$.
To put this planet  mass in context to mass scales for planet--disc interaction, the thermal mass scale from \citet{2002ApJ...572..566R}  is
\begin{align}
M_1 &=\frac{2 c_{\rm s}^3}{3 \Omega G} = 1.4\times10^{-4} \ ,
\end{align}
and the Toomre Q parameter of the disc at the planet's initial location is
\begin{align}
Q &= \frac{\Omega c_{\rm s}}{\pi G \Sigma} = 17.5 \ ,
\end{align}
which combined yields the disc feedback mass from \citet{2002ApJ...572..566R} as
\begin{align}
M_{\rm F} &\simeq 3.8\left(\frac{Q}{h}\right)^{-5/13} M_1=  5.5\times 10^{-5} \ ,
\end{align}
so the planet mass used here is approximately half this mass, which in 2D isothermal discs marks a transition to
a regime of disc feedback modified migration in inviscid discs.
Thus the planet is firmly in the embedded regime, and well below the feedback mass.
In the understanding derived from two-dimensional calculations \citep{2019MNRAS.484..728M},
we thus expect the planet to be in the regime where migration behaviour is significantly influenced by
by dynamical corotation torques \citep{2014MNRAS.444.2031P,2017MNRAS.472.1565M,2018MNRAS.477.4596M}, as discussed in Section~\ref{sec:theory}.

The planet potential is smoothed by taking the planet to be a sphere with uniform density and radius of $0.1$ scale heights.
A torque cut-off within the Hill sphere is employed in calculating the torques, to eliminate spurious contributions
originating from poorly resolved parts of the flow.
Hence, the torque from the disc on the planet from each grid cell is tapered by the function $\xi(r_{\rm c})$:
\begin{align}
\xi(r_{\rm c}) & = \begin{cases}
0 &\text{ if } r_{\rm c}/\rH < 1/2\\
\left(\sin(r_{\rm c}/\rH-1/2)\right)^2 &\text{ if } 1/2 \leq r_{\rm c}/\rH \leq 1\\
1 &\text{ if } r_{\rm c}/\rH > 1 \ ,
\end{cases}\\
\rH & =  (q/3)^{1/3} \rp \ ,
\end{align}
where $r_{\rm c}$ is the distance from the cell centre to the planet position, and $\rH$ is the Hill radius.
At higher resolutions it may be possible to avoid this compromise in the future.
Note, different planet potential smoothing and torque calculations are used in 2D comparisons, as described below.

When torque values are shown we non-dimensionalize them in terms of $\Gamma_0$
\begin{align}
\Gamma_0 \equiv \left(\frac{q}{h}\right)^2 \Sigma_{\rm p} r_{\rm p}^4 \Omega_{\rm p}^2\ ,
\end{align}
where $\Sigma_{\rm p}$ is the surface density at the planet's position in the background disc model,
and $\Omega_{\rm p}$ is the planet's orbital angular velocity.

The planet is held fixed on a Keplerian orbit at $\rp=2$ for the first 200 orbits, after which it is released
and follows the torque exerted by the gas, including the correction for the neglect of self-gravity of \citet{2008ApJ...678..483B}.
Additionally, the planet mass is ramped linearly from zero to the final planet mass over the same period.
In comparison to these time-scales, an estimate of the libration time for gas on the outermost librating orbit of the corotation region can be derived
from an estimate of the corotation region width.
In two-dimensional approximation from \citet{2010MNRAS.401.1950P} assuming $b/h=0.4$, where $b$ is the softening length, the corotation region half-width is
\begin{align}
\xs &= \frac{1.1}{\gamma^{1/4}} \left(\frac{0.4}{b/h}\right)^{1/4}\sqrt{\frac{q}{h}} \rp = 0.0405\ ,
\end{align}
which gives a libration time of
\begin{align}
\tau_{\rm lib} &= \frac{8\pi \rp}{3 \Omegap \xs} = 1171\ ,
\end{align}
or 65 orbits at $r=2$.

\subsection{Two-dimensional comparison simulations}
The two-dimensional simulations are constructed along the lines of ones conventionally used
for disc--planet interactions, derived from vertically integrating Equations~\ref{eq:continuity}--\ref{eq:eos}.
We follow the convention of using an equation of state with the same numerical value of $\gamma$ in two dimensions as in the
three-dimensional models.
Because these 2D simulations are a vertically integrated model of the disc,
  the planet potential must be changed
to a Plummer sphere potential \citep{2012A&A...546A..99K}.
For the same consistency reasons, no Hill sphere tapering of the disc torques is employed, 
as this scale is smaller than the disc scale height for a low-mass planet.
We specifically employ a smoothing length for the Plummer potential of
 $b=0.4h$ for consistency with our previous 2D works.
As wind driving of surface layers is inherently 3D, only models without wind driving and adiabatic thermodynamics were run.

\subsection{Analysis of Vortensity}
The vorticity-weighted surface density of the corotation region is central to
 the theory of horseshoe torques in 2D \citep{1991LPI....22.1463W,1992LPI....23.1491W,2009MNRAS.394.2283P,2014MNRAS.444.2031P}.
In the large mass regime where the planet modifies the disc surface density, the vortensity is dominated by the mass-weighting 
leading to the concept of a coorbital mass deficit in runaway or Type-III migration \citep{2003ApJ...588..494M,2007prpl.conf..655P}.
 To extend this analysis of horseshoe torques to three dimensions, \citet{2016ApJ...817...19M}
 showed that the equivalent quantity that can be derived from a three-dimensional model is a scalar:
\begin{align}
V = \left[\int_{-\infty}^{+\infty}\left(\frac{\zeta_z}{\rho}\right)^{-1} dz\right]^{-1} \label{eq:vertintvort},
\end{align}
where $\zeta_z$ is the vertical component of the vorticity ${\bm \zeta}$, which in the rotating frame is given by
\begin{align}
{\bm\zeta} = \nabla \times \V + 2\Omega_{\rm F}.
\end{align}
For simplicity, we will refer to this quantity $V$ as the (vertically integrated) vortensity.
Note that this quantity is distinct from the three dimensional potential vorticity, which would be a vector quantity ${\bm \zeta}/\rho$.

In 2D, the vortensity is a scalar quantity as ${\bm \zeta} = \zeta {\bm \hat{e}}_{z}$ and the potential vorticity and vortensity are identical.
Thus, in two-dimensional cases, we refer to $V=\zeta/\Sigma$ as vortensity.

\section{Results}
\label{sec:results}

We begin the presentation of simulation results with two-dimensional models, which generally yield results
in agreement with previous simulations and theoretical expectations.
We then proceed to discuss the analogous three-dimensional models, which display new and surprising behaviours.

\subsection{Two-dimensional models}
\label{sec:twodmodels}

In 2D models, the planet trajectories, once released, show only small oscillations due to the presence of the vortices discussed in Section~\ref{sec:theory},
and agreement between the planet trajectories in single and double resolution simulations is excellent (see Fig.~\ref{fig:trajectories_2d3d}).
The torque histories in Figure~\ref{fig:torques_2AMS_2AMD_2AFS} show that although the planet trajectory and time-averaged torque agrees well between the single and double resolution simulations,
the instantaneous oscillations driven by vortices in the double-resolution case are much larger.

Comparison of the torque history in the moving planet cases to the
fixed planet case in Figure~\ref{fig:torques_2AMS_2AMD_2AFS} agrees with expectations from
 previously established dynamical corotation torque concepts.
The time-averaged torque in the fixed planet case 2AFS agrees well with the formula given by \citet{2010MNRAS.401.1950P}, 
their equation 14,
 for the Lindblad torque alone,
which indicates that the corotation torque is saturated.
Both the estimates of the Lindblad plus linear unsaturated corotation torque and the Lindblad plus full unsaturated horseshoe torques from \citet{2010MNRAS.401.1950P} 
(their equations 18 and 46 respectively)
are significantly above
the torque observed in run 2AFS, further suggesting that the corotation torque is fully saturated.
From the theory of dynamical corotation torques discussed in Section~\ref{sec:theory}, a positive corotation toque is expected to arise
for an inward moving planet in this disc \citep{2014MNRAS.444.2031P}, and the slowing of the inward migration, combined with the reduction of the magnitudes of the torques, for the runs 2AMS and 2AMD are in agreement with this expectation.

Maps of the vortensity in an annulus near the planet are shown in Fig.~\ref{fig:V_map_2AMS_2AMD_2AFS}
for the moving and fixed planet cases, along with the double-resolution moving planet case.
These will later be contrasted to maps from the equivalent three-dimensional models shown in Fig.~\ref{fig:V_map_3AFS_3AMS_3AFS}
In the moving planet case, the planet brings with it low-vortensity material from the outer disc trapped on librating streamlines.
The contrast is moderated by instabilities at the edge of the libration island, which are particularly well resolved at the higher resolution in run 2AMD
 \citep{2001MNRAS.326..833B,2014MNRAS.444.2031P}.
However, when the planet is held fixed on one orbit, the libration island is merely a
well-mixed average of the disc background at this radial location (Fig.~\ref{fig:V_map_2AMS_2AMD_2AFS}, lower panel 2AFS).
This contrast in the depletion of  vortensity is particularly clear in the azimuthally averaged
vortensity relative perturbation shown in  Fig.~\ref{fig:Vavg_2AFS_2AMS}.
This depletion, combined with the geometrical asymmetry of the flow on U-turn trajectories
having close encounters with the planet in front and behind it azimuthally give rise to the positive dynamical corotation torque.
This simple agreement with the theory of \citet{2014MNRAS.444.2031P} for dynamical corotation toques (see Section~\ref{sec:theory}) is in contrast to the
new effects observed in three dimensions, as discussed below.

In these two-dimensional simulations, abundant vortices can be seen being spawned from the sharp vortensity contrast at
the edge of the libration island, particularly at the exit side of the U-turns in front and behind of the planet.
In the radial direction, these newly formed vortices  are typically much smaller than the gas scale height (0.1 at $r=2$)
and they are initially only a few scale heights in the azimuthal direction.
However, in two-dimensional flows, interacting vortices have a tendency to merge and grow to larger vortices.
The results of this are particularly clear in the high-resolution run 2AMD in Fig.~\ref{fig:V_map_2AMS_2AMD_2AFS}
where both small primary vortices and large merged vortices can be seen in the libration island, and long-lived small
vortices persist in the disc flow downstream (radially outside) of the planet.
The tighter, denser vortex cores in the double-resolution run 2AMD corresponds to the larger
instantaneous oscillations in the torque compared to the single resolution run 2AMS.

\subsection{Three-dimensional models}
\label{sec:threedmodels}

The trajectories of the moving planets in the purely adiabatic and wind+thermal relaxation runs 3AMS and 3WMS are
shown with the analogous 2D simulation 2AMS  in Fig.~\ref{fig:trajectories_2d3d}.
It is immediately apparent that in 3D, the planets migrate inwards more rapidly than in 2D, and that the wind
driving model does not produce a significantly different outcome\footnote{Comparing the green lines in Figs~\ref{fig:torques_2AMS_2AMD_2AFS} and \ref{fig:torques_3AMS_3WMS_3AFS} shows there is a small offset in the Lindblad torques associated with the 2D and 3D runs. This offset wants to drive more rapid migration in the 2D runs compared to the 3D runs. The fact that the opposite is observed is due to the differing evolution of the corotation torques.}.

This conclusion concerning the inefficacy of the wind driving in altering the torque is held up by
the torque histories of these simulations shown in Fig.~\ref{fig:torques_3AMS_3WMS_3AFS}.
However, the surprising aspect of that figure is the sign of the offset between the time-averaged torque
in the fixed planet case and the moving planet cases --- the moving planets have a more negative torque than
the fixed planet, the opposite of that observed in 2D (Fig.~\ref{fig:torques_2AMS_2AMD_2AFS}).
This negative dynamical torque effect has the opposite sign from that expected by previous dynamical corotation torque theory
\citep{2014MNRAS.444.2031P}.
A comparison of these results to a half-resolution simulation is presented in Appendix~\ref{sec:3dresolution}.

To search for a possible origin of this extra negative torque effect, the time-averaged azimuthal surface density perturbations
for the moving and fixed planet cases are presented in Fig.~\ref{fig:coldensdiff}.
A relative surface density enhancement appears behind  the planet in its orbit in the moving planet case,
while the surface density distribution is more symmetrical in the fixed planet case.
This corresponds to an extra mass dragging backwards on the planet in its orbit in the moving planet case,
or a negative dynamical corotation torque.
Considering the dynamical corotation torque theory of \citet{2014MNRAS.444.2031P} one would expect this mass enhancement to arise from
a more spatially concentrated vortensity enhancement.

Maps of the vortensity in an annulus near the planet in the three runs are shown in
Fig.~\ref{fig:V_map_3AFS_3AMS_3AFS}.
There is a strong contrast to the two-dimensional case shown in Fig.~\ref{fig:V_map_2AMS_2AMD_2AFS},
in that a vortensity enhancement is clearly present in the coorbital region, whereas in 2D a deficit was observed.
Three dimensionality also significantly modifies the appearance and breakdown of vortices,
and diagonal features in the vertically averaged vortensity appear, due to the vertically plunging flows near 
the planet and the buoyancy resonance (explored in Section \ref{sec:buoyancy}).
Most surprisingly, in the fixed planet case 3AFS (Fig.~\ref{fig:V_map_2AMS_2AMD_2AFS}, lower panel),
the vortensity of the libration island is still enhanced,
although because of the lack of radial movement of the planet the libration island is azimuthally symmetrical and hence does not give rise to a corotation torque.
The azimuthally averaged vertically integrated vortensity shown in Fig.~\ref{fig:Vavg_2AFS_3AFS}
for the fixed planet case indeed shows that in 3D, even for a fixed planet, the vortensity in the coorbital region becomes enhanced
over the background value, while no enhancement is observed in the equivalent 2D model.
This enhancement, combined with the geometrical asymmetry of the flow on U-turn trajectories
having close encounters with the planet in front and behind it azimuthally give rise to the negative
dynamical corotation torque \citep{2014MNRAS.444.2031P},
the opposite of the effect observed in 2D.

In these 3D models, like in 2D, vortices are spawned at the exit of the U-turns in front and behind the planet.
In the moving planet cases, the U-turn in front of the planet dominates.
The diagonal stripe vortensity features are also strongest in this region,
and so make distinguishing vortices difficult.
However, as the vortices proceed to be advected around the edge of the libration island,
they tend to weaken or break up, as opposed to merging into larger structures.
The primary vortices, like in 2D, are typically a fraction (about one quarter)
of a scale height in extent in the radial direction, and a few scale heights in the azimuthal direction.

\subsection{Passively irradiated disc models}
\label{sec:irrad}

\begin{figure}
\includegraphics[width=0.9\columnwidth]{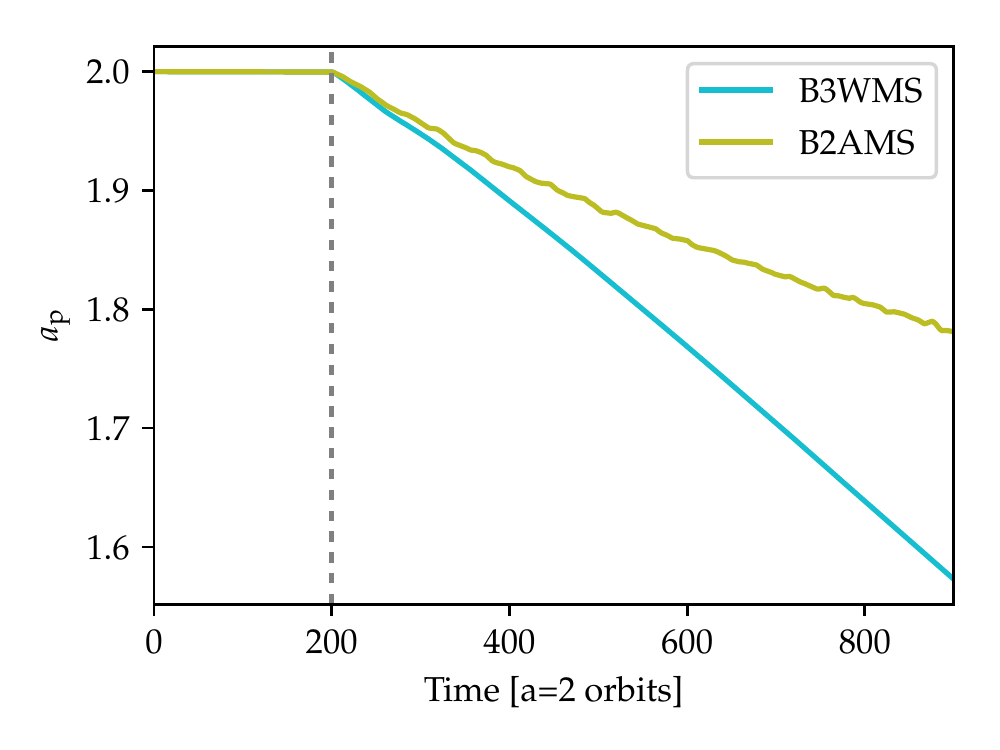}\\
\caption{
Planet migration trajectories in 2D and 3D models with $h\propto r^{2/7}$. {\sl Grey dashed line:} Planet release time.
}
\label{fig:trajectories_irr_2d3d}
\end{figure}

\begin{figure*}
\includegraphics[width=1.8\columnwidth]{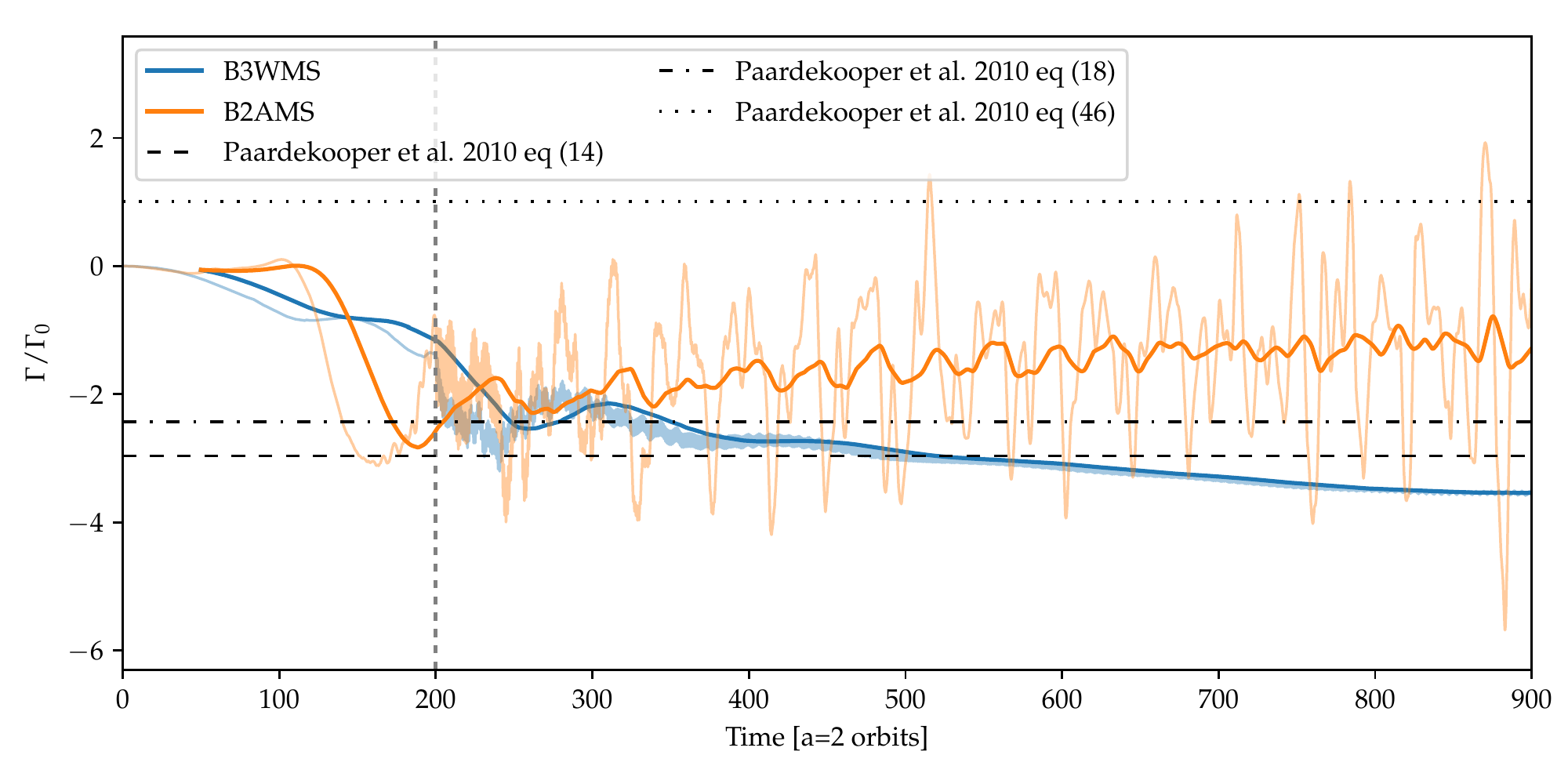}\\
\caption{
Torque histories from models with irradiated disc temperature profiles.
Model B3WMS: Irradiated disc 3D wind plus thermal relaxation prescription, moving planet, single resolution.
Model B2AMS: Irradiated disc 2D adiabatic equation of state moving planet single resolution.
{\sl Light lines:} Instantaneous values.
{\sl Solid lines:} 50 orbit trailing averages.
{\sl Grey dashed vertical line:} End of planet mass ramping and planet release.
Also plotted are 2D torque formulas from \citet{2010MNRAS.401.1950P}. Their equation~(14) is an analytical estimate of the Lindblad torque, 
equation~(18) is the Lindblad torque and linear corotation torque, and equation~(46) the Lindblad torque and unsaturated non-linear horseshoe corotation torque.
}
\label{fig:torques_B3WMS_B2AMS}
\end{figure*}

\begin{figure}
\includegraphics[width=\columnwidth]{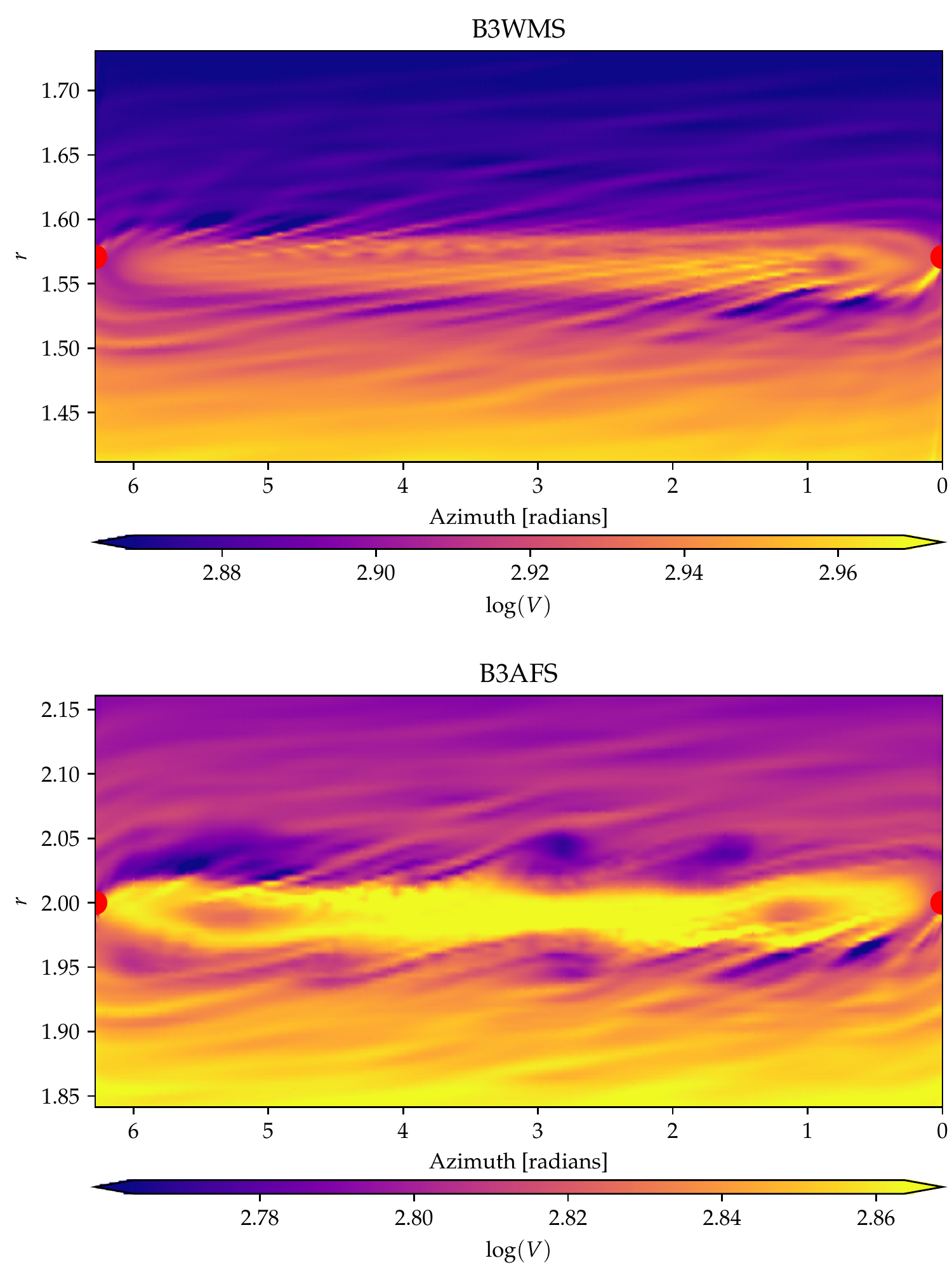} \\
\caption{
Vertically integrated vortensity $V$ in the runs B3WMS (at 900 orbits) and B3AFS (at 325 orbits), corresponding to a passively irradiated disc,
 in a strip centred on the planet location.
The colour scales are a fixed logarithmic range about the vortensity of the initial condition at the planet's present location.
The planet location marked by the red circle split across on the periodic azimuthal boundary.
}
\label{fig:V_map_B3WMS_B3AFS}
\end{figure}

In Sections~\ref{sec:twodmodels}--\ref{sec:threedmodels} we presented
models of discs with a constant aspect ratio.
However, a wind-driven disc is likely to have a radial temperature profile corresponding to a passively irradiated thermal equilibrium.
The radial temperature scaling of such a disc should be $T\propto r^{-3/7}$ with
the radial scaling of the aspect ratio $h\propto r^{2/7}$ \citep{1997ApJ...490..368C}.
In this section we discuss equivalent simulations to the constant aspect ratio runs, but with this different radial temperature
power law.
 These are referred to as B3WMS (analogous to run 3WMS with a wind, thermal relaxation, and moving planet)
 and B3AFS (analogous to run 3AFS with adiabatic gas and a fixed planet).

 Similar behaviour with respect to the action of dynamical corotation torques between 2D and 3D
 models are seen here.
 Moving planets migrate faster in 3D than in 2D (Fig.~\ref{fig:trajectories_irr_2d3d}).
 In 2D, a positive dynamical corotation torque  retards the inward migration, whereas in 3D
 a negative corotation torque enhances the inward migration over the pure
 Lindblad torque rate (Fig.~\ref{fig:torques_B3WMS_B2AMS}).
 Comparing Fig.~\ref{fig:torques_B3WMS_B2AMS} and the torque from the equivalent 3D run 3WMS in
 Fig.~\ref{fig:torques_3AMS_3WMS_3AFS}, it is notable that the torque for
 B3WMS lies clearly below the \citet{2010MNRAS.401.1950P} Lindblad torque estimate.
 This makes the role of the negative dynamical corotation torque more clear in the B3WMS case.

Like in the constant aspect ratio case, the vertically integrated vortensity $V$ increases in the libration
island in both the moving and fixed planet cases, as shown in Fig.~\ref{fig:V_map_B3WMS_B3AFS}.
It is notable that in the moving planet case B3WMS, the widespread vortices present in all other 3D cases are absent.
This result suggests that those vortices are not required to drive the increasing vortensity of the libration island.

\subsection{Buoyancy response}
\label{sec:buoyancy}

\begin{figure*}
\includegraphics[width=1.8\columnwidth]{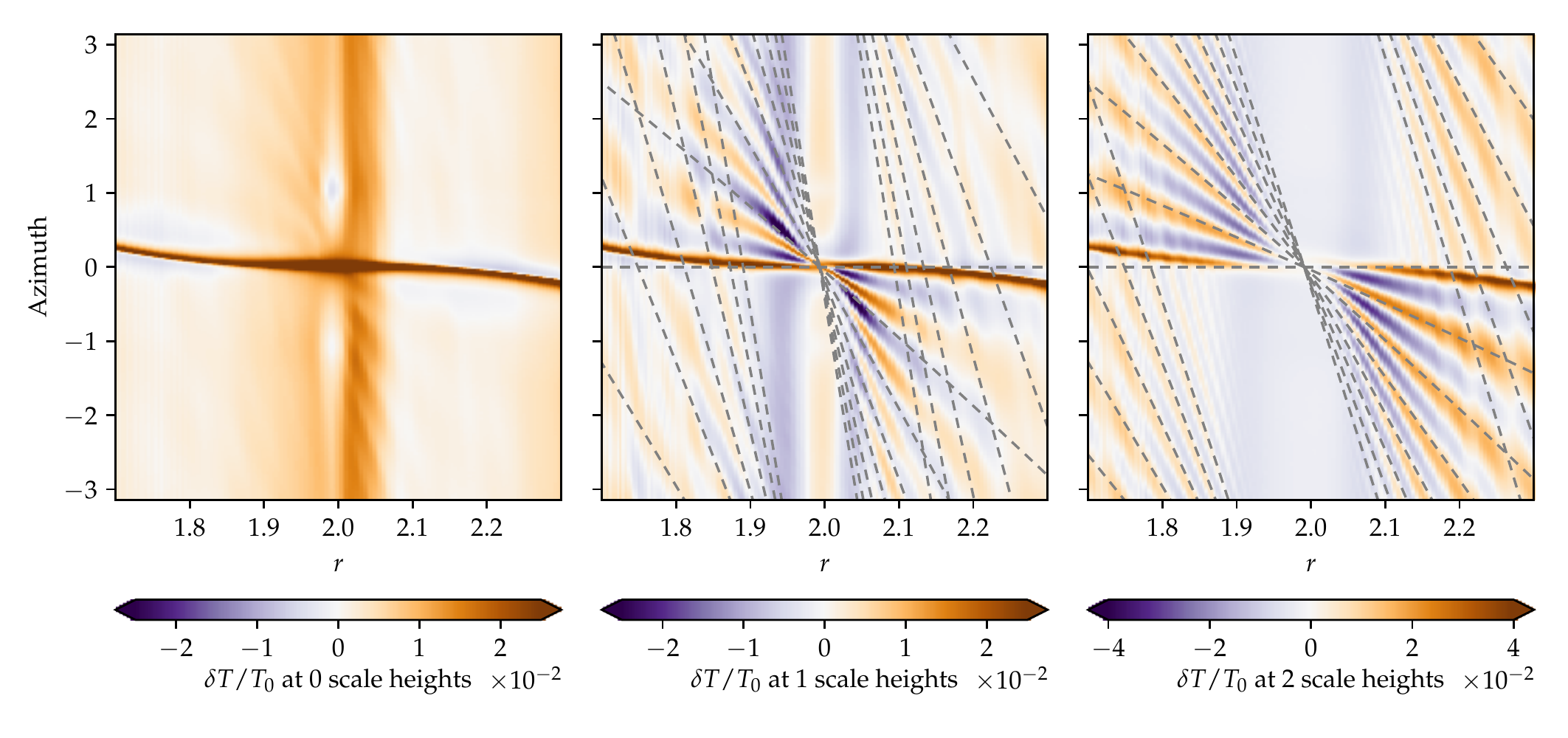}
\caption{
Temperature perturbations at constant polar angle averaged from 200 to 500 orbits, disc with vertically isothermal equilibrium structure (run 3AFS).
{\sl Left:} Mid-plane
{\sl Centre:} 1 scale height
{\sl Right:} 2 scale heights
{\sl Grey dashed lines:} Positions of constant phase of linear buoyancy response.
}
\label{fig:wavelines_temperature_3AFSI}
\end{figure*}

\begin{figure*}
\includegraphics[width=1.8\columnwidth]{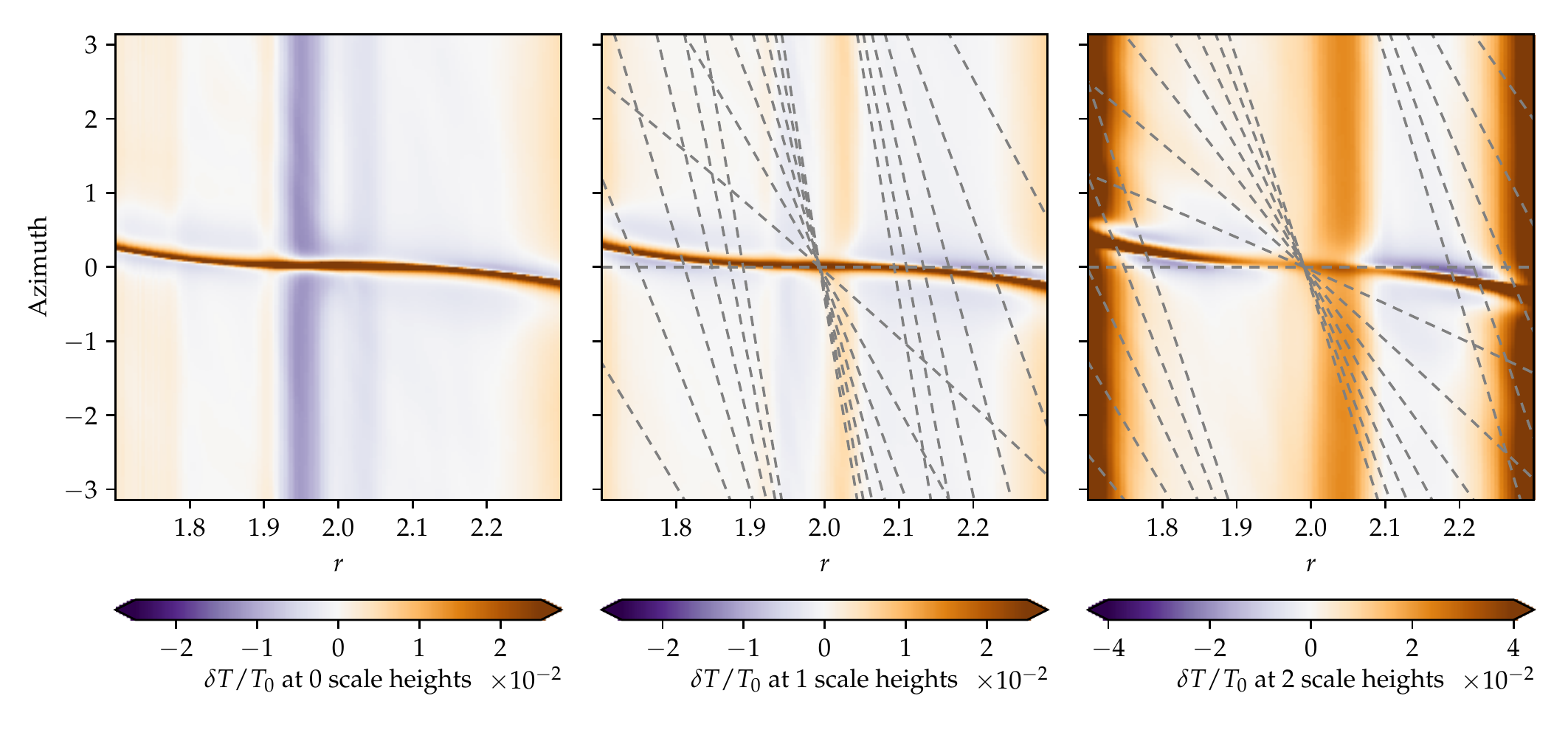}
\caption{
Temperature perturbations at constant polar angle averaged from 200 to 500 orbits, disc with polytropic equilibrium structure (run 3AFSP).
{\sl Left:} Mid-plane
{\sl Centre:} 1 scale height
{\sl Right:} 2 scale heights
{\sl Grey dashed lines:} Positions of constant phase of linear buoyancy response.
Note the lack of features following the dashed grey lines giving the predicted locations of buoyancy waves
as present in the disc with vertically isothermal equilibrium structure in Fig.~\ref{fig:wavelines_temperature_3AFSI}.
}
\label{fig:wavelines_temperature_3AFSP}
\end{figure*}

\begin{figure}
\begin{center}
\includegraphics[width=0.9\columnwidth]{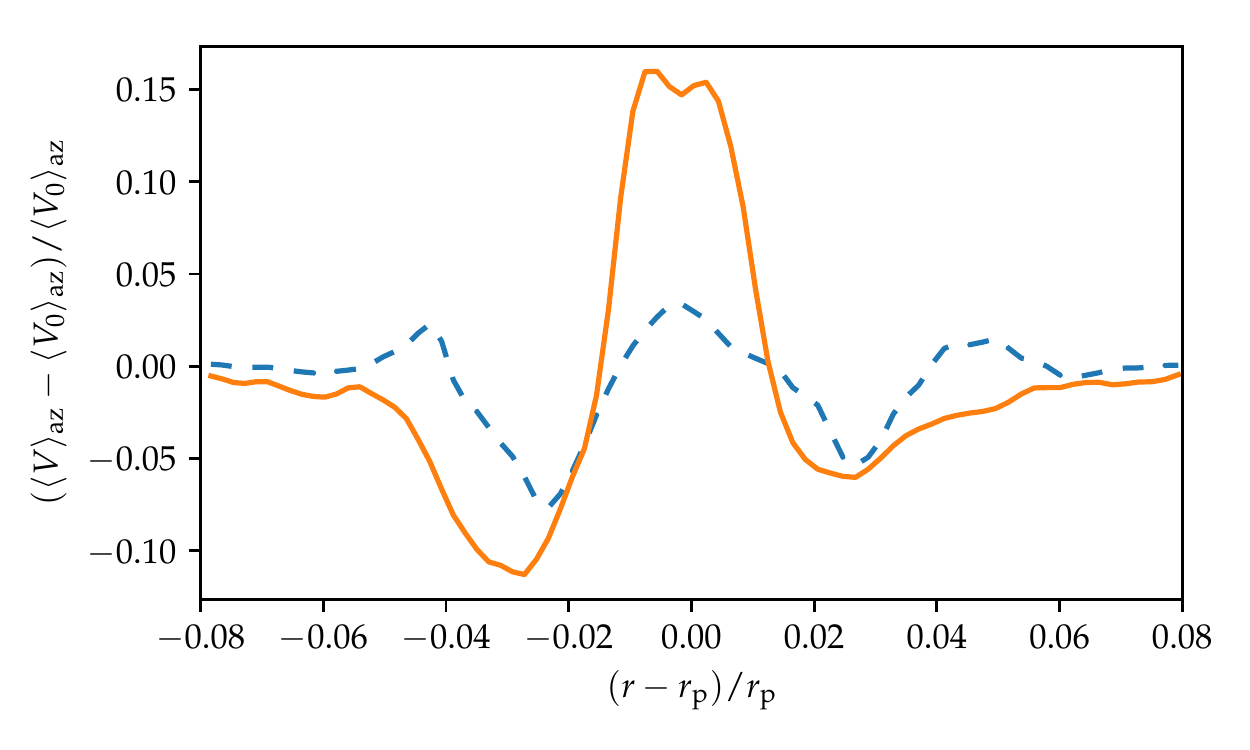} \\
\end{center}
\caption{
Azimuthally averaged  vertically integrated vortensity relative perturbation, at 500 orbits.
{\sl Solid line:} Three-dimensional model 3AFS with isothermal vertical structure ($\Gamma=1$) and adiabatic gas dynamics ($\gamma=1.4$).
{\sl Dashed line:} Three-dimensional model 3AFSP with polytropic vertical structure ($\Gamma=1.4$) matching the adiabatic index of the gas ($\gamma=1.4$).
This demonstrates how the vortensity enhancement in the coorbital region relies on the non-zero \BV frequency.
}
\label{fig:Vavg_3AFSI_3AFSP}
\end{figure}

\begin{figure}
\begin{center}
\includegraphics[width=\columnwidth]{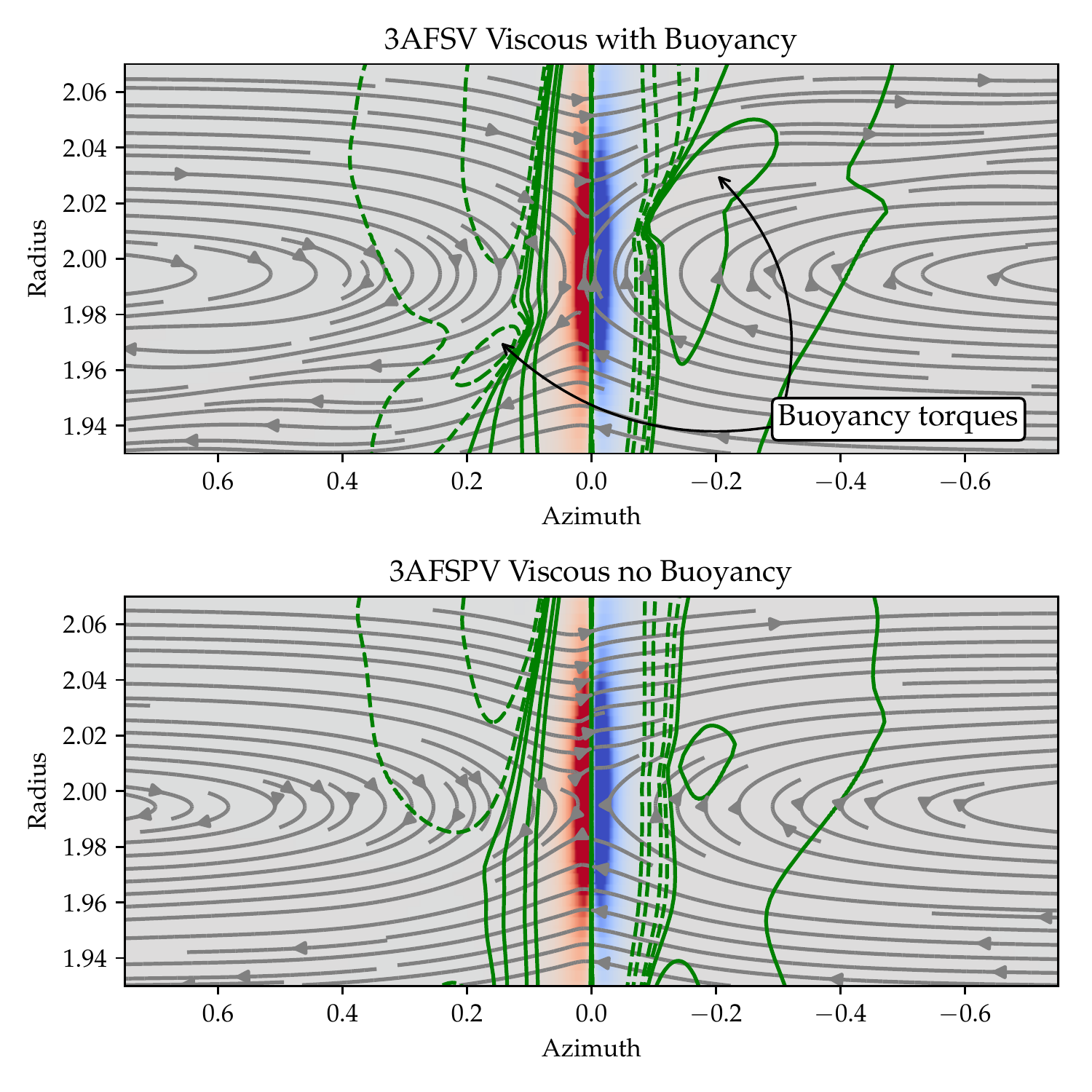} \\
\end{center}
\caption{
Comparison of torque density in a disc with and without the presence of the buoyancy response,
with a fixed planet and a small kinematic viscosity.
{\sl Upper:} Run 3AFSV
{\sl Lower:} Run 3AFSPV
{\sl Color:} Total vertically integrated torque density acting on planet, red positive, blue negative.
{\sl Contours:} Relative torque density change with respect to the initial condition density field, solid lines indicate positive contours, dashed lines indicate negative contours.
{\sl Streamlines:} Gas velocity in the frame of the planet at the disc mid-plane.
In the upper panel, the regions corresponding to the buoyancy torques are annotated.
}
\label{fig:paperplots/visctorq_122_123_comp}
\end{figure}

\begin{figure}
\begin{center}
\includegraphics[width=\columnwidth]{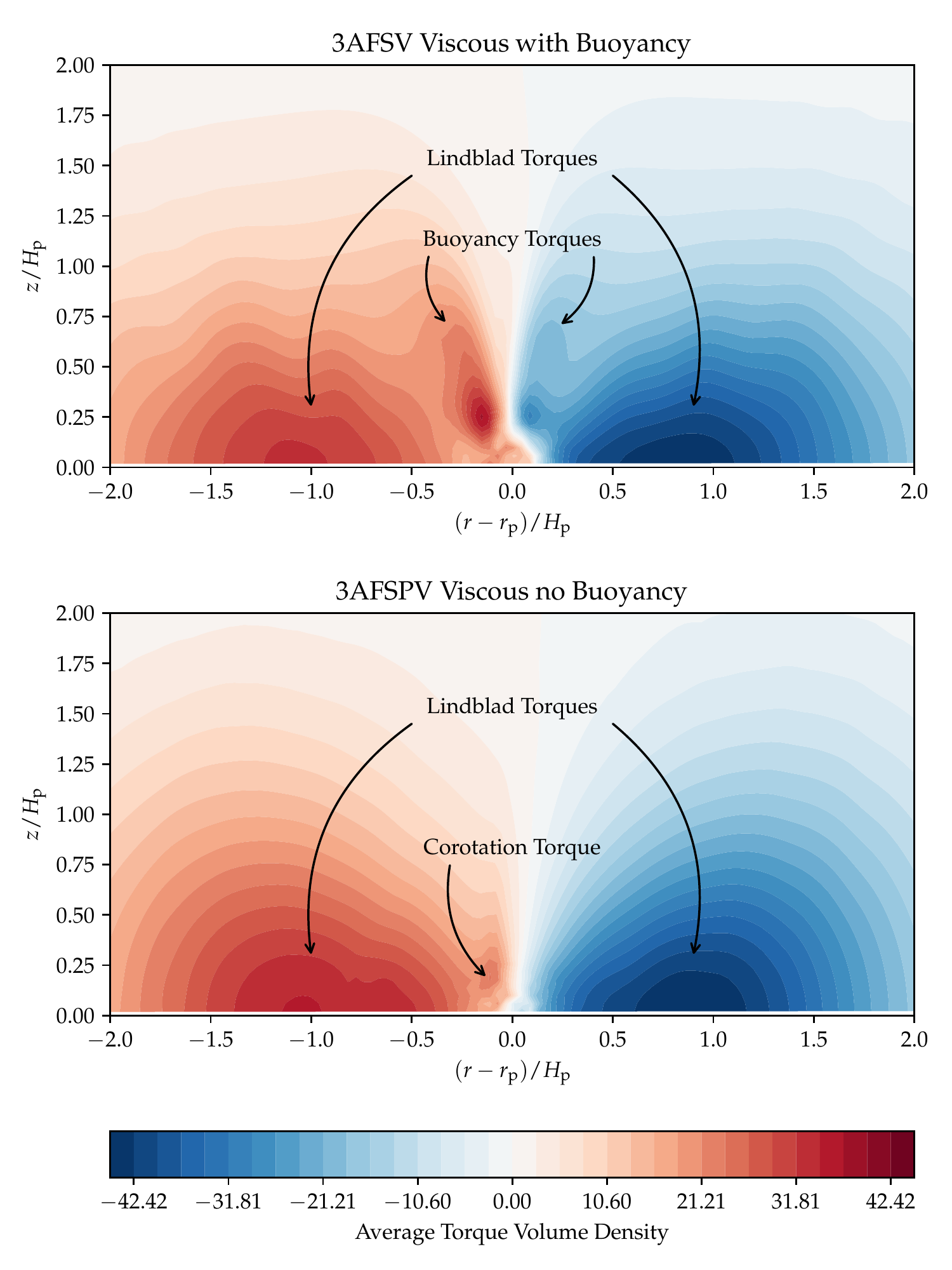} \\
\end{center}
\caption{
Comparison of torque density in a disc with and without the presence of the buoyancy response,
The planet is located at the origin $(0,0)$.
Annotations point out the lobes formed by the Lindblad torques, buoyancy torques, and the corotation torque.
}
\label{fig:paperplots/torquecut_3AVSIV_3AFSPV}
\end{figure}

\begin{figure}
\begin{center}
\includegraphics[width=\columnwidth]{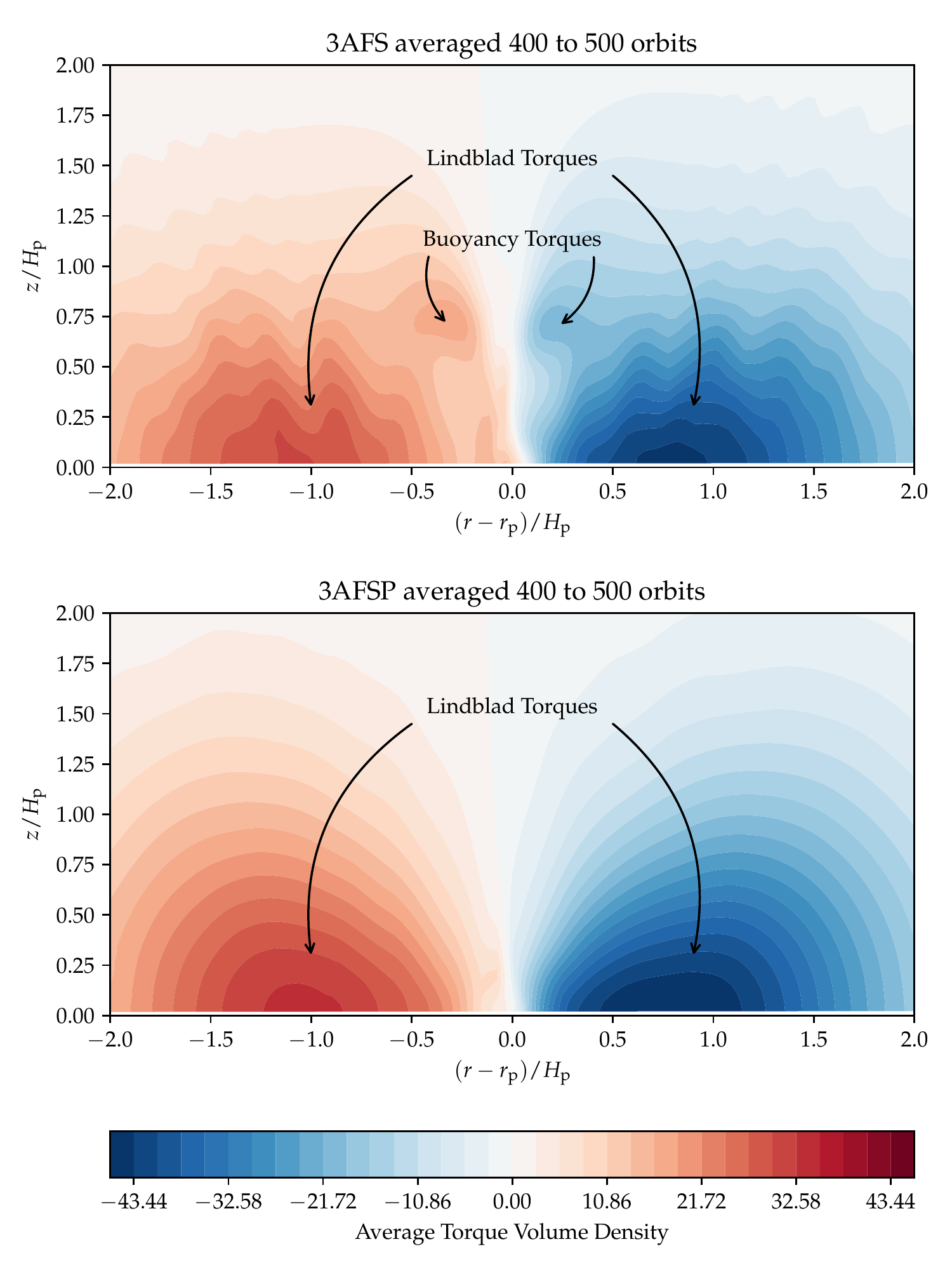} \\
\end{center}
\caption{
Comparison of azimuthally and temporally averaged torque density in discs with
and without the presence of the buoyancy response.
The planet is located at the origin $(0,0)$.
}
\label{fig:paperplots/torquecut_3AVSI_3AFSP}
\end{figure}

Gas parcels moving past the planet, above the mid-plane, are accelerated towards the mid-plane by the planet's gravity.
Buoyancy responses in the disc oscillate at the \BV frequency, which in the vertical-radial meridional plane is
\begin{align}
N^2 = N^2_r + N^2_z = -\frac{1}{\rho c_{\rm p}} \nabla P \cdot \nabla s \ ,
\end{align}
where $N_r$ and $N_z$ are pieces of the \BV frequency corresponding to the frequency for purely vertical and radial oscillations
and $c_{\rm p}$ is the gas specific heat capacity at constant pressure \citep{2019PASP..131g2001L}.
Here, following  \citet{2015ApJ...813...88Z} we will analyze the oscillations purely in terms of the \BV frequency for vertical oscillations $N_z$, which is
\begin{align}
N^2_z =  -\frac{1}{\rho c_{\rm p}} \frac{\partial P}{\partial z} \frac{\partial s }{\partial z} \ .
\end{align}
If the \BV (buoyancy) frequency is positive, these fluid elements will oscillate vertically
around their equilibrium position in the disc atmosphere.
This is the essential difference between 2D and 3D disc models.
As given by \citet{2015ApJ...813...88Z}, assuming zero phase at the planet position $\phi=0$, the lines of constant phase of
vertical oscillations due to the buoyancy response are given by
\begin{align}
\phi= -2 n \pi (\Omega_{\rm p} - \Omega) \sqrt{\frac{\gamma}{\gamma-1}} \frac{H}{\Omega_{\rm K} z}\left( 1 + \frac{z^2}{R^2}\right)^{3/2}
\label{eq:buoyancy_theta}
\end{align}
for $n=0, 1, 2, \ldots$ where $z$ is the height above the mid-plane and $R$ the cylindrical radius.
The angular velocities in this expression are $\Omega$, the angular velocity of the gas, and $\Omega_{\rm K}$, the Keplerian angular velocity.
Two sets of effects that may lead the positions of the buoyancy resonances to be radially asymmetric with
 respect to the planet position can be readily identified.
First, the planet orbits slightly faster than the gas at its radial location, meaning that the buoyancy
resonance locations, determined by the relative motion of
the planet and gas, are shifted in radius to be asymmetrical with respect the the planet.
Second, the radial gradients in the vertical scale height, density, and thermal structure of the disc
cause radial gradients of the \BV frequency at a given height above the mid-plane.

The influence of the buoyancy response on the vertically integrated vortensity of the corotation region can be largely removed by
using a disc vertical structure where the \BV frequency goes to zero \citep{2015ApJ...813...88Z}.
That is, a disc where the hydrostatic structure obeys $P\propto \rho^\Gamma$ with $\Gamma=\gamma$ the
adiabatic index of the gas equation of state.
The required polytropic global equilibrium disc structure is given in \citet{2013MNRAS.435.2610N}, their equations (14--15). We have run such a model, labelled as 3AFSP, to denote its vertically polytropic nature, to compare with the previously discussed vertically isothermal run 3AFS.

%
%

In the model with vertically isothermal structure 3AFS, and a varying but positive \BV frequency,
the temperature fluctuations in the gas, averaged over the interval $200$--$500$ orbits, shows clearly
in Fig.~(\ref{fig:wavelines_temperature_3AFSI}) where the buoyancy response produces rays in the flow downstream of the planet.
The positions of these rays matches the predicted pattern for the buoyancy response from equation~\ref{eq:buoyancy_theta}.
In contrast, in the disc with vertically polytropic structure, 3AFSP, these same patterns are absent as shown in Fig.~(\ref{fig:wavelines_temperature_3AFSP}).
This is because the gas does not oscillate vertically in a buoyant manner once displaced in the polytropic disc structure.

The long-term consequence of this difference in the presence or absence of the buoyancy response in 3D is analogous to the difference
between the evolution of vortensity in the corotation region of 2D and 3D models shown in Fig.~\ref{fig:Vavg_2AFS_3AFS}.
In this 3D case, the buoyancy response drives an enhancement of the vortensity in the corotation region in the case of a disc with isothermal vertical structure,
whereas this enhancement is absent in the case with a vertically polytropic structure as shown in Fig.~\ref{fig:Vavg_3AFSI_3AFSP}.

\subsubsection{Torque contour maps with buoyancy torques}
\label{sec:torquemapbuoyancy}

The primary challenge with analysis of the buoyancy torque is determining the sign of the net effect.
Buoyancy-resonances on the inside of the planet's orbit exert a positive torque on the planet, and to the outside a negative torque
 \citep{2014ApJ...785...32L}.
To determine the sign of the net torque from simulation outputs would require isolating the
 buoyancy-related torque from Lindblad and corotation torques.
In this section we present analysis showing how the buoyancy related
 torque is spatially overlaid on the Lindblad and corotation torques.
We have not found it possible to disentangle these contributions in a simple and clear manner.
Nevertheless, these results provide a clearer graphical description of buoyancy torque related
structures than have appeared before in the literature.

To damp vortex formation and produce a time-steady flow, we ran
viscous versions of  the runs 3AFS and 3AFSP, with $\nu=10^{-7}$, labelled here as 3AFSV and 3AFSPV.
While producing a time-steady flow which can be instantaneously analysed,
the viscosity has the important side effect of preventing a build-up of vortensity in the libration region
and partially unsaturating the classical corotation torque.
None the less, we present analysis of the torque distributions in these models as
they provide useful interpretative context for the inviscid models.

A comparison of the vertically integrated torque spatial distributions and mid-plane gas streamlines is shown in
Fig.~\ref{fig:paperplots/visctorq_122_123_comp}.
The presence of new features in the torque distribution due to the buoyancy torque is clear, but subtle.
We remind the reader that the contours shown are of the relative change of the torque density with respect to the initial condition (smooth disc).
This visualization, as opposed to contours of the total torque density itself, brings out the buoyancy torque features, but does make interpretation difficult.
These 
buoyancy-related features
appear as diagonally extended contours in the residual torque (torque density in evolved state minus torque density in the background state),
aligned with the beginning of the radial wobble that may be observed in the mid-plane flow streamlines, driven by the buoyancy resonance downstream of the planet.

The spatial arrangement of the Lindblad and buoyancy torques is slightly more clear in the radial-vertical
plane showing the azimuthally averaged torque density.
In Fig.~\ref{fig:paperplots/torquecut_3AVSIV_3AFSPV}
the torque distribution is dominated by two large lobes comprising the Lindblad torque,
centred at the mid-plane roughly one scale height away from the planet on each side.
In the vertically isothermal run, 3AFSV, an additional pair of positive and negative torque lobes are
also present at a height $z/H_{\rm p}\approx 0.25$ just radially interior and exterior to the planet's location.
These lobes have a tilt away from the vertical axis running through the planet location, corresponding to
the shift of the buoyancy resonance location with height.
However, in the corresponding run with vertically polytropic structure, 3AFSPV, the pair of buoyancy torque lobes are absent,
and only a single near-vertical positive signed structure just interior to the planet's orbit is present,
indicative of the partially unsaturated classical corotation torque.

The equivalent plot from time-averaging the torque distribution over 100 orbits
of inviscid models, to smooth the effects of vortices,
is shown in Fig.~\ref{fig:paperplots/torquecut_3AVSI_3AFSP}.
Here, in the vertically isothermal structure case, 3AFS, vertical striations
tilted slightly out away from the planet's location are present, at least five on each side.
By comparison to the vertically polytropic case, 3AFSP, these appear to be the result of the buoyancy resonance.
The most prominent feature in the run 3AFS is the inner pair of buoyancy torque lobes,
with peaks at a height $z/H_{\rm p}~\sim 0.75$ above the mid-plane.
These are notably asymmetrical, with the inner one slightly higher and radially further from the planet.

\section{Discussion}
\label{sec:discussion}

\begin{figure}
\includegraphics[width=\columnwidth]{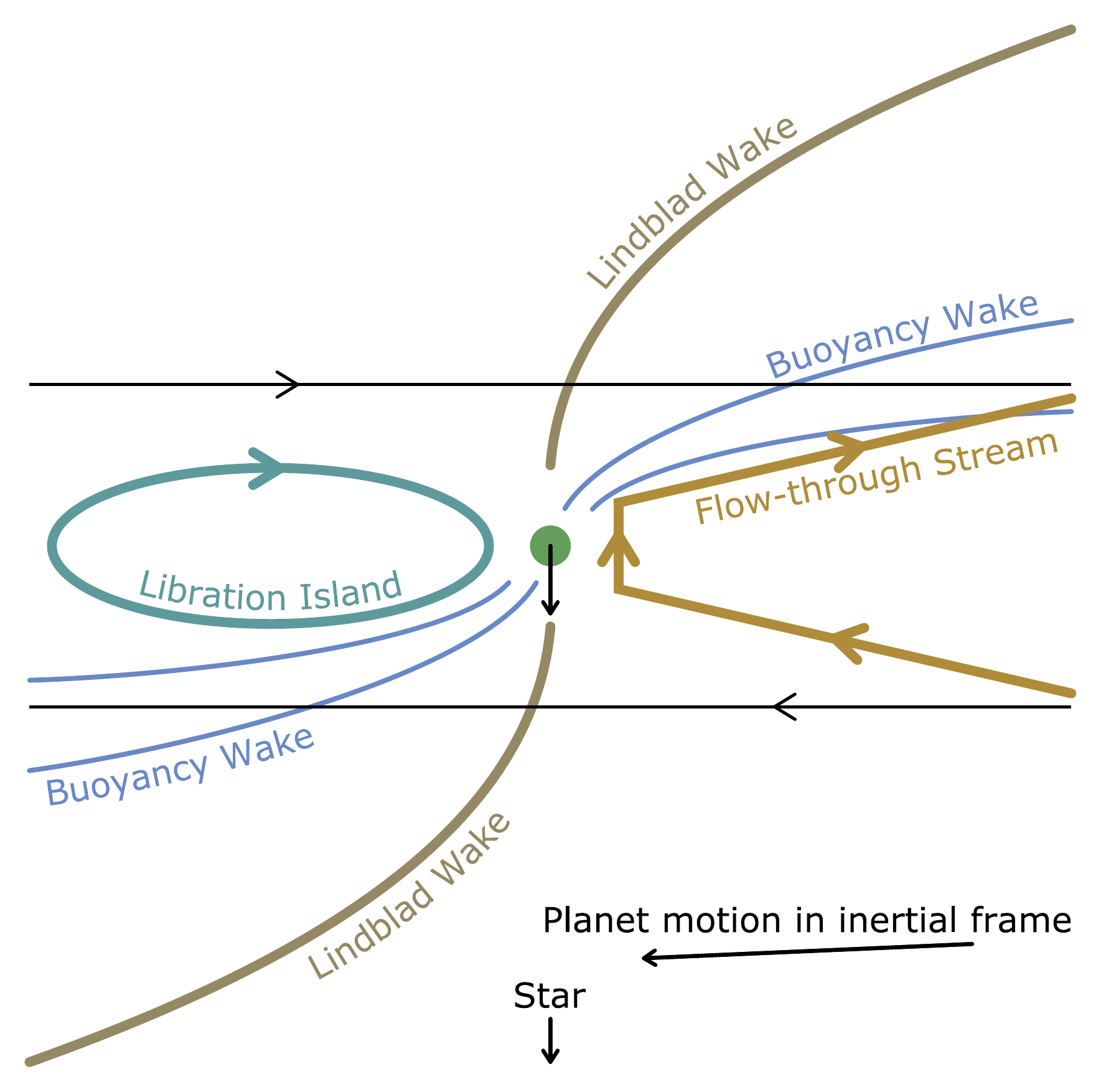} \\
\caption{
Features driving angular momentum exchanges for a planet migrating inwards through a disc.
Lindblad torques arise from the over densities in the Lindblad wake.
The response of the disc at the buoyancy resonance produces the buoyancy wake (with multiple features), and the buoyancy torque.
In the corotation region, the material trapped in the libration island and material passing across the planet's
orbit in the flow-through stream both exchange angular momentum with the planet, producing the corotation torque.
}
\label{fig:planet_disc_diagram}
\end{figure}

Our results point to a new and important phenomenon when considering the migration of low-mass planets in discs which sustain very low levels of turbulent viscosity, namely that the dynamical corotation torque evolves in a qualitatively different manner in 3D compared to expectations based on earlier 2D calculations.
The fundamental expectation for the evolution of the torque on a planet moving through an
 inviscid disc established by \citet{2006MNRAS.370..784O} was that the torque will depend only on two dimensionless parameters:
 a diffusion time-scale and a radial speed.
 In \citet{2017MNRAS.472.1565M} we established that the relevant radial speed can be considered as not just
 the radial speed of the planet with respect to the central star, but the relative speed with respect to the background disc.
 In three dimensions, our simulations show that this parameterisation does not capture the full dependence,
 and the rate at which the action of the buoyancy response can alter the vortensity of
  the libration island must also be a parameter controlling the problem. 
 Of more fundamental importance than this, however, is that our results show that the sign of the corotation torque can be changed by the buoyancy response.
 Although in 2D the slowing and eventual stalling of migration was to be expected by the dynamical corotation torque, in 3D the corotation torque acts to accelerate inward migration. 
 Hence, the de facto expectation that low-mass planet migration may not occur over large distances at the rapid rate induced by the Lindblad torque, because of the slowing effect of the dynamical corotation torque, now needs to be revised.

 \subsection{Angular momentum exchanges}
 Our simulations point to a complex array of mechanisms by which angular momentum is exchanged between a low-mass planet and an inviscid disc, leading eventually to a well-defined net migration torque. Here, we attempt to outline how these various exchange processes operate.
 \begin{description}
 \item[\bf Lindblad torques] Spiral density waves are launched at Lindblad resonances, which are located away from the planet's coorbital region, and the waves propagate into the disc away from the coorbital region carrying with them an angular momentum flux.
 The density waves that propagate into the inner/outer disc provide a negative/positive torque on the disc, and the angular momentum the waves carry is deposited in disc material where the waves eventually dissipate.
This occurs through non-linear steepening in inviscid discs \citep{Goodman2001}, and the waves produce the Lindblad wake (Fig.~\ref{fig:planet_disc_diagram}).
Conservation of angular momentum means that spiral waves launched at outer Lindblad resonances exert a negative torque on the planet, with the opposite being true for waves at inner Lindblad resonances. The outer Lindblad resonances are stronger than the inner ones, so a planet experiences a net negative torque, while the disc receives angular momentum from the planet.
 \item[\bf Buoyancy torques] Unlike Lindblad resonances, buoyancy resonances are present close to the planet within the coorbital region.
The disc response produces the buoyancy wake (Fig.~\ref{fig:planet_disc_diagram}).
 Previous work on the buoyancy response of a disc to the presence of a planet indicates that the one-sided buoyancy torque provides a contribution that increases the effects of the one-sided Lindblad torque by $\sim 10\%$ \citep{2012ApJ...758L..42Z}, 
 although we caution that this value must be dependent on the planet and disc properties. 
 Hence, the inner/outer buoyancy resonances lead to the disc being negatively/positively torqued. To date there has not been any work that determines the magnitude of the differential buoyancy torque arising from summing the effects of inner and outer buoyancy resonances, and hence it is uncertain what the instantaneous net buoyancy torque acting on a planet ought to be. \citet{2014ApJ...785...32L} undertook an analytic study of the buoyancy response, and concluded that the disc responds in a non-wave like manner at buoyancy resonances. This suggests that a large fraction of the torque exerted by the planet at buoyancy resonances should be deposited locally in disc material. Our simulations for planets on fixed orbits demonstrate unequivocally that the corotation region is subject to a net negative torque arising from the two-sided buoyancy response, because the observed increase in  vortensity can only occur when the corotation region is negatively torqued \citep{2017MNRAS.472.1565M}. While this does not prove that the total buoyancy torque exerted on the disc by the planet is negative, since angular momentum may be deposited across a range of radii in the disc, it does show that the inner buoyancy resonances are more effective at torquing disc material in the corotation region than their outer disc counterparts.
 \item[\bf Corotation torques] The dynamical corotation torque acting on a planet depends on the net angular momentum exchange occurring when gas undergoes horseshoe u-turns in front of and behind the planet (Fig.~\ref{fig:planet_disc_diagram}). The angular momentum associated with the material undergoing u-turns is related to its vortensity. When the planet migrates inwards, some of the material that makes a u-turn behind the planet originates from the background disc outside of the corotation region. This gas has a single encounter with the planet as it is effectively scattered from the inner to the outer disc, giving rise to a negative torque on the planet. The gas that undergoes u-turns in front of the planet consists entirely of trapped coorbital material that is on librating horseshoe streamlines. This material exerts a positive torque on the planet whose magnitude depends on its vortensity. When the vortensity of the corotating material is larger than that in the background disc then the net corotation torque is negative (see equation~\ref{eq:unified}), and an inward migrating planet migrates faster due to the corotation torque. Hence, the effect of a buoyancy torque removing angular momentum from the coorbital region, causing the vortensity to increase there, is to change the balance of angular momentum exchange at the two u-turns such that the material flowing directly from the inner to outer disc provides the dominant angular momentum exchange leading to a negative corotation torque.
 \end{description}

 \subsection{Influence of an accretion flow}
 Another important result unveiled by our simulations is the fact that a laminar accretion flow located only in the disc surface layers, due to a putative magnetocentrifugal wind being launched there, does not influence the migration of a low-mass embedded planet. In earlier work \citep{2017MNRAS.472.1565M, 2018MNRAS.477.4596M}, we showed that a laminar accretion flow located at the disc mid-plane would have a strong influence on migration. For a gas inflow rate slower than the migration induced by the Lindblad torque alone, the evolution of corotation torque would slow the planet's migration and asymptotically it would migrate at the same speed as the disc gas. For fast gas inflow, the corotation torque would eventually cause the planet's migration to reverse and move outwards in a runaway fashion. In the simulations presented here, the gas flow rate in the disc surface layers corresponds to fast flow, and we observe that the planet maintains inward migration at a rate similar to that of a planet in a disc without a surface accretion flow. Hence, it would seem that there is a minimum depth to which the surface accretion flow must penetrate before noticeable effects occur, and these are unlikely to be realized in a realistic protoplanetary disc.

The situation may be different, however, when a vertical magnetic field threads the disc and is aligned with the rotation vector. The Hall effect may then lead to the formation of strong horizontal magnetic fields near the mid-plane that wind up and induce a laminar radial gas flow, an effect that does not occur if the field and rotation vector are oppositely aligned \citep{2017ApJ...845...75B,2017A&A...600A..75B}. This is essentially the scenario explored using a 2D approximation in the above cited papers \citep{2017MNRAS.472.1565M, 2018MNRAS.477.4596M}, and although we have not explored this effect in 3D in this paper, we have no reason to believe that the 2D results are not robust. This is one way in which the influence of the buoyancy response of the disc on the corotation torque could be counteracted. For a slow radial accretion flow, planet migration would still be accelerated by the evolution of the corotation torque because the buoyancy response of the disc causes the vortensity of the coorbital disc material to evolve too quickly. A fast radial accretion flow, however, would cause the migration to slow down, stop and eventually reverse.

\subsection{Caveats and future work}
The \BV frequency, which determines the behaviour of buoyancy waves and the locations of buoyancy resonances,
 depends on the disc's thermal structure and evolution.
 In this work, we considered vertically isothermal models, and adopted a simple thermal relaxation approach to modelling the entropy evolution of the gas.
While the isothermal approximation may be justified below the Rosseland mean photosphere, where
the disc is optically thick to its own thermal radiation, the upper, optically thin regions
of the disc will have a much higher temperature \citep{1997ApJ...490..368C}.
This however, is likely of little consequence when the disc has a significant migration-driving surface density,
as such regions occur only high in the disc column where the contribution to
the torques and net angular momentum flows are small. What is likely to be important, however, for accurately capturing the buoyancy response of the disc is treating the internal energy transport in a realistic fashion. This will be one area of focus for future work.

A related issue is the possibility of local heating
of the gas near the planet at the mid-plane that can occur as a result of the accretion
luminosity of an embedded planet.
This heating can itself alter the disc thermal and density structure
sufficiently to give rise to a heating torque \citep{2015Natur.520...63B}.
The primary effects are seen close to the planet \citep{2017MNRAS.472.4204M},
but this heating process may
 be an additional effect which can alter the buoyancy response,
  and hence also the dynamical corotation torque, when present.

Ideally, in three dimensions the flow down to the planetary surface would be resolved,
to include the density structure and thermodynamics of the most compressed gas.
Similar to heating torques \citep{2015Natur.520...63B}, our
simulations also do not produce a `cold finger' effect of a torque contribution from
cold gas streams drawn close to the planet by the u-turn flow
\citep{2014MNRAS.440..683L}.
Although it has not been demonstrated in inviscid simulations, future models
will need to include the thermal diffusion required to produce the effect.
In the case that heating torques only affect the flow very close to the planet,
the combinations of heating torques and cold finger-like effects can lead to
overdensities near the planet moving in a way to drive oscillations in
the planet migration \citep{2019A&A...626A.109C}.
By analogy to vortices passing near the planet in the u-turn
 flows seen in models like 3AMS, we would not expect these oscillations
 to annul the effects of the buoyancy response on the libration
  island and the subsequent dynamical corotation torque.
However, it is clear that an important goal of further 3D simulations should be to include all the required thermodynamics,
to assess if a scale separation between heating torque effects local to
the planet and dynamical corotation torque effects in the
 libration island and flow-through stream can be made, and to determine whether or not the effects
 reported here are robust to a fuller treatment of the thermal history of the gas.

In the regime in which this study is posed, the Ohmic resistivity is so overwhelmingly large in the mid plane layers of the disc
that even if the Hall effect was able to generate toroidal fields, they would be effectively 
decoupled from the planet-induced flows \citep{ 2017MNRAS.472.1565M}.
In the case that the disc body is torqued by a spiral magnetic field, such that the disc can be described as an `advective disc' 
\citep{2017MNRAS.472.1565M,2018MNRAS.477.4596M,2019MNRAS.484..728M, 2018haex.bookE.139N},
the principle of the Galilean relativity of the dynamical corotation torque \citep{2017MNRAS.472.1565M}
should apply in three dimensions as it does in two.
As the planet mass increases, and the planet-induced gap in the disc becomes deeper, a transition to a regime where higher ionization allows 
 magnetic field effects to become more relevant to the migration torque should occur.
If the planet is situated either radially inwards of the Ohmic dead zone, or far enough radially outwards in the disc, 
mid-plane coupling of the magnetic field may again influence the corotation torque directly
 \citep{2003MNRAS.341.1157T,2005MNRAS.363..943F,2011A&A...533A..84B,2013MNRAS.430.1764G}.

In this work, we have only studied models with a single planet mass and
disc surface density due to the computational expense of these simulations.
There must be a dependence of the buoyancy and vortex effects
on the planet mass, disc surface density, and optical depth of the disc.
Although many more expensive three-dimensional simulations will surely be required,
an analytical understanding of the unknown process by which the buoyancy response is
able to modify the libration island would be an important guide.

\section{Conclusions}
\label{sec:conclusions}
We have presented the highest resolution and longest evolving inviscid low-mass planet--disc interaction simulations in 3D to date.
These also include the first test of low-mass planet--disc interaction with a model for wind-driven accretion layers located at disc surfaces.

It is well known that in 2D inviscid disc models, inward migration of a planet through the disc leads to the build up of a dynamical corotation torque that can slow or even stall the migration \citep{2014MNRAS.444.2031P}. This arises because of the geometrical asymmetry of the librating horseshoe region for a migrating planet, combined with the evolution of the vortensity of the trapped librating material relative to that of the background disc gas. 
In three dimensions, we have found that the ability of disc gas to respond through vertical motions to the planet potential
at buoyancy resonances leads to evolution of the vortensity of the libration region,
even without relative radial motion of the planet and disc, and this vortensity evolution has the dramatic effect of changing the sign of the dynamical corotation torque. 
This torque acts in addition to a direct buoyancy torque arising from the buoyancy resonance \citep{2012ApJ...758L..42Z,2014ApJ...785...32L}.
Consequently, the dynamical corotation torque in 3D models accelerates the inward migration of a low-mass planet instead of slowing it down.

Our previous work has shown that a laminar accretion flow located near the mid-plane of a protoplanetary disc can have a strong influence on the migration of a low-mass planet \citep{2017MNRAS.472.1565M,2018MNRAS.477.4596M}. In the case of a rapid inward gas flow, where the in-flow speed is faster than the speed at which the planet naturally wants to migrate, the planet's inward migration can be completely reversed. The 3D simulations we have presented which include surface accretion flows, adopted fast in-flow rates to examine the effects of accretion confined to disc surface layers on planet migration. We find that because the accreting layers are located at high altitudes in the disc, where only a small fraction of the gas resides, the influence of these accreting layers on migration is essentially negligible. There does not appear to be any communication between the torqued accreting surface layers and the passive mid-plane regions, which would lead to modification of the planet migration.

In both 2D and 3D, with an ideal-gas equation of state and adiabatic or slow cooling gas,
vortices, generated baroclinically by the mixing of entropy across the horseshoe region, infect the corotation region of low-mass planets at low viscosity.
In 2D and 3D we find the vortex evolution differs markedly, as expected.
In 2D models, small vortices formed below the scale height merge and grow, whereas in 3D a
stronger tendency for these narrow vortex structures to break down to smaller vortices and dissipate is observed. In both 2D and 3D, these vortices appear to play a role in mixing material between the librating horseshoe region and the background disc, and hence regulate the contrast that develops in the vortensity of the librating material versus that in the background disc, influencing the rate at which dynamical corotation torques develop.

Our simulations adopted a number of simplifying assumptions regarding the thermal evolution of the gas, and this may influence the ability of buoyancy resonances to drive the evolution of the corotation region, and hence modify the dynamical corotation torques. In future work we will  examine the migration of low-mass planets under a more refined treatment of the thermodynamic evolution to test the robustness of the results we have presented here.

\section*{Acknowledgements}
We thank the referee for a constructive and detailed report.
This research was supported by STFC Consolidated grants awarded to the
 QMUL Astronomy Unit 2015--2018  ST/M001202/1 and 2017--2020 ST/P000592/1.
We acknowledge that the results of this research have been achieved using the PRACE Research Infrastructure resource Irene based in France at TGCC.
This research utilised Queen Mary's Apocrita HPC facility, supported by QMUL Research-IT \citep{apocrita}
 (whom we are particularly grateful to for a loan of 100~TB of storage);
the DiRAC Data Centric system at Durham University, operated by the Institute for Computational Cosmology on behalf
of the STFC DiRAC HPC Facility (www.dirac.ac.uk).
This equipment was funded by a BIS National E-infrastructure capital grant ST/K00042X/1, STFC capital grant ST/K00087X/1,
DiRAC Operations grant ST/K003267/1 and Durham University.
DiRAC is part of the National E-Infrastructure.
SJP is supported by a Royal Society University Research Fellowship.
This project has received funding from the European Union's Horizon
2020 research and innovation programme under grant agreement No 748544
(PBLL).
The research leading to these results has received funding from the European Research Council under the European Union's Horizon 2020 research and innovation programme (OG, grant agreement No 638596).




\bibliographystyle{mnras}
\bibliography{threeddisc_paper} 


\appendix
\section{3D resolution study}
\label{sec:3dresolution}
\begin{figure*}
\includegraphics[width=1.8\columnwidth]{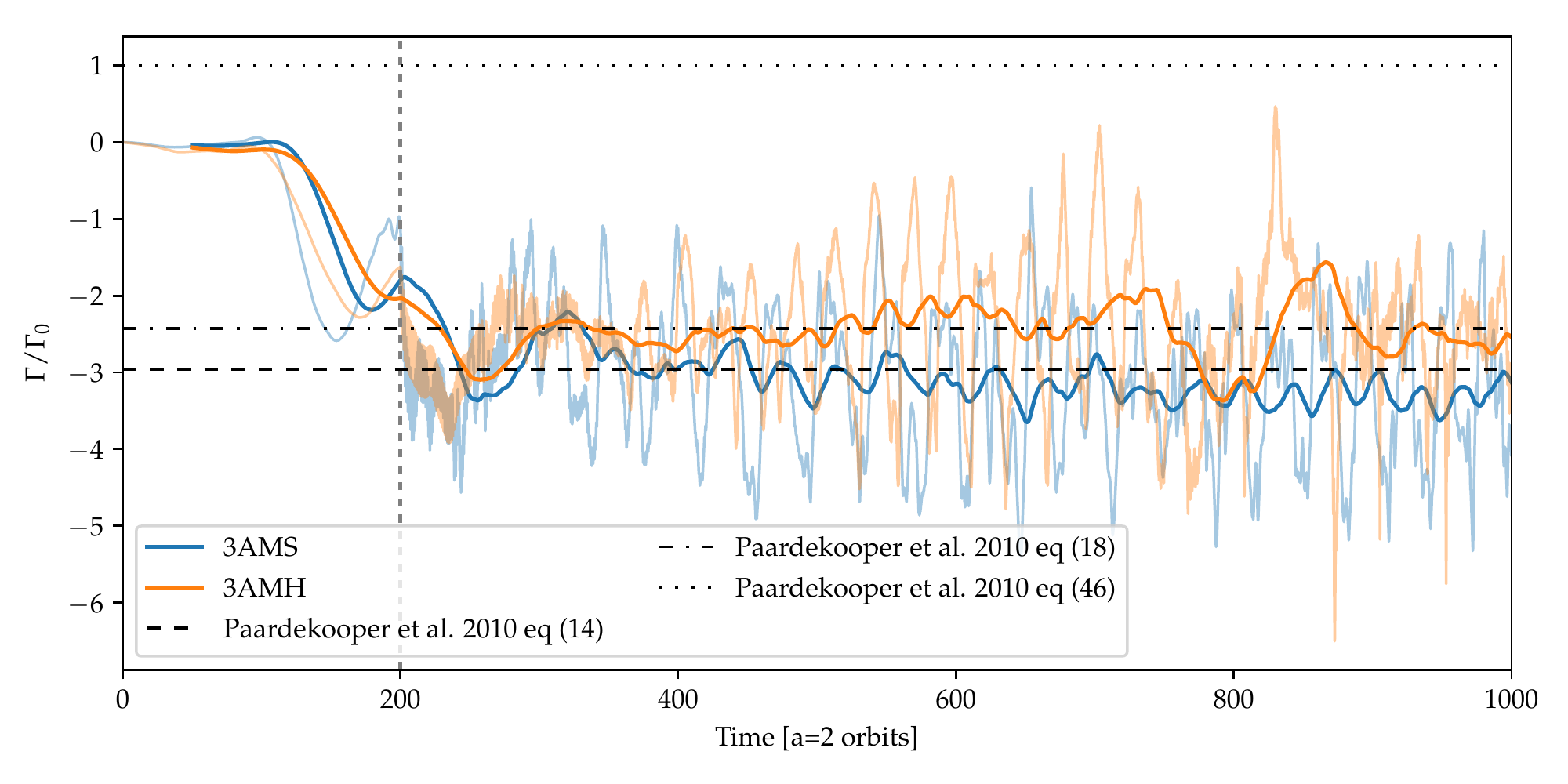}\\
\caption{
Torque histories from 3D resolution study of moving planet adiabatic thermodynamics case.
Run names given in Table~\ref{tab:runs}.
{\sl Light lines:} Instantaneous values.
{\sl Solid lines:} 50 orbit trailing averages.
{\sl Grey dashed vertical line:} End of planet mass ramping and planet release.
}
\label{fig:torques_3AMS_3AMH}
\end{figure*}
In the three-dimensional models, a higher resolution was not computationally feasible, so a study at lower
resolution was performed. Fig.~\ref{fig:torques_3AMS_3AMH} presents the single-resolution run 3AMS and the equivalent run with
half the spatial resolution 3AMH.
At this lower resolution the dynamical corotation effect arising in the very geometrically thin corotation region is significantly reduced.
This suggests the possibility that even in our highest resolution simulation, 3AMS, the
full magnitude of the dynamical corotation torque effect is not yet realized.
However, as the flow in this case is not laminar and has significant mixing of the librating region due to vortices, it
is plausible that these Implicit Large-Eddy simulations will converge due to the effect of the largest
 vortices mixing vortensity out of the librating material.

\bsp	
\label{lastpage}
\end{document}